\documentclass[12pt,english]{article}
\usepackage[T1]{fontenc}
\usepackage[latin9]{inputenc}
\usepackage{color}
\usepackage{amsthm}
\usepackage{amsmath}
\usepackage{amssymb}
\usepackage{babel}
\usepackage[matrix,frame,arrow]{xy}
\usepackage{cite}

  \theoremstyle{plain}
  \newtheorem{lem}{Lemma}
  \theoremstyle{plain}
  \newtheorem*{lem*}{Lemma}
\theoremstyle{plain}
\newtheorem{thm}{Theorem}
\theoremstyle{plain}
  \newtheorem{prop}{Proposition}

\usepackage{amsfonts}
\usepackage{amsthm}
\usepackage[labelfont=bf, font=small]{caption}

\usepackage{xy}
\xyoption{matrix}
\xyoption{frame}
\xyoption{arrow}
\xyoption{arc}

\begin{document}
\global\long\global\long\def\kb#1#2{|#1\rangle\langle#2|}
\global\long\global\long\def\bk#1#2{\langle#1|#2\rangle}
\global\long\global\long\def\ket#1{|#1\rangle}
\global\long\global\long\def\bra#1{\langle#1|}
\global\long\global\long\def\c#1{\mathbb{C}^{#1}}

\newcommand{\linespace}{\vspace{\baselineskip}}
\newcommand{\semilinespace}{\vspace{0.5\baselineskip}}
\newcommand{\upline}{\vspace{-\baselineskip}}
\newcommand{\upupline}{\vspace{-\baselineskip}\vspace{-\abovedisplayskip}}
\newcommand{\be}{\begin{equation}}
\newcommand{\ee}{\end{equation}}
\renewcommand{\(}{\left(}
\renewcommand{\)}{\right)}
\newcommand{\pr}{^\prime}
\newcommand{\g}{\mathfrak{g}}
\newcommand{\hh}{\mathfrak{h}}
\newcommand{\ssll}{\mathfrak{sl}}
\newcommand{\Lie}{\mathfrak}
\newcommand{\K}{\mathbb{K}}
\renewcommand{\a}{\mathbf{a}}
\renewcommand{\b}{\mathbf{b}}
\renewcommand{\c}{\mathbf{c}}
\renewcommand{\u}{\mathbf{u}}
\renewcommand{\v}{\mathbf{v}}
\newcommand{\w}{\mathbf{w}}
\newcommand{\x}{\mathbf{x}}
\newcommand{\y}{\mathbf{y}}
\newcommand{\z}{\mathbf{z}}
\newcommand{\del}{\nabla}
\newcommand{\A}{\mathbf{A}}
\newcommand{\B}{\mathbf{B}}
\newcommand{\CC}{\mathbf{C}}
\newcommand{\LL}{\mathbf{L}}
\renewcommand{\P}{\mathbf{P}}
\newcommand{\Q}{\mathbf{Q}}
\newcommand{\RR}{\mathbf{R}}
\renewcommand{\S}{\mathbf{S}}
\newcommand{\X}{\mathbf{X}}
\newcommand{\operone}{\mathbf{1}}
\newcommand{\R}{\mathbb{R}}
\newcommand{\C}{\mathbb{C}}
\renewcommand{\O}{\mathbb{O}}
\newcommand{\F}{\mathcal{F}}
\renewcommand{\L}{\mathcal{L}}
\newcommand{\M}{\mathcal{M}}
\newcommand{\ox}{\otimes}
\newcommand{\ol}{\overline}
\newcommand{\<}{\langle}
\renewcommand{\>}{\rangle}
\newcommand{\half}{\tfrac{1}{2}}
\newcommand{\hlf}[1]{\tfrac{#1}{2}}
\newcommand{\third}{\tfrac{1}{3}}
\newcommand{\thrd}[1]{\tfrac{#1}{3}}
\newcommand{\quarter}{\tfrac{1}{4}}
\newcommand{\eighth}{\tfrac{1}{8}}
\newcommand{\qqquad}{\quad\quad\quad}
\newcommand{\qqqquad}{\quad\quad\quad\quad}
\newcommand{\imp}{\Rightarrow}
\newcommand{\impl}{\Longrightarrow}
\newcommand{\ad}{\operatorname{ad}}
\newcommand{\id}{\operatorname{id}}
\newcommand{\rad}{\operatorname{rad}}
\newcommand{\End}{\operatorname{End}}
\newcommand{\tr}{\operatorname{tr}}
\newcommand{\transpose}{^{\mathrm{T}}}
\renewcommand{\Re}{\operatorname{Re}}
\renewcommand{\Im}{\operatorname{Im}}

\newcommand{\qw}[1][-1]{\ar @{-} [0,#1]}
\newcommand{\qwx}[1][-1]{\ar @{-} [#1,0]}
\newcommand{\cw}[1][-1]{\ar @{=} [0,#1]}
\newcommand{\cwx}[1][-1]{\ar @{=} [#1,0]}
\newcommand{\gate}[1]{*+<.6em>{#1} \POS ="i","i"+UR;"i"+UL **\dir{-};"i"+DL **\dir{-};"i"+DR **\dir{-};"i"+UR **\dir{-},"i" \qw}
\newcommand{\meter}{*=<1.8em,1.4em>{\xy ="j","j"-<.778em,.322em>;{"j"+<.778em,-.322em> \ellipse ur,_{}},"j"-<0em,.4em>;p+<.5em,.9em> **\dir{-},"j"+<2.2em,2.2em>*{},"j"-<2.2em,2.2em>*{} \endxy} \POS ="i","i"+UR;"i"+UL **\dir{-};"i"+DL **\dir{-};"i"+DR **\dir{-};"i"+UR **\dir{-},"i" \qw}
\newcommand{\measure}[1]{*+[F-:<.9em>]{#1} \qw}
\newcommand{\measuretab}[1]{*{\xy*+<.6em>{#1}="e";"e"+UL;"e"+UR **\dir{-};"e"+DR **\dir{-};"e"+DL **\dir{-};"e"+LC-<.5em,0em> **\dir{-};"e"+UL **\dir{-} \endxy} \qw}
\newcommand{\measureD}[1]{*{\xy*+=<0em,.1em>{#1}="e";"e"+UR+<0em,.25em>;"e"+UL+<-.5em,.25em> **\dir{-};"e"+DL+<-.5em,-.25em> **\dir{-};"e"+DR+<0em,-.25em> **\dir{-};{"e"+UR+<0em,.25em>\ellipse^{}};"e"+C:,+(0,1)*{} \endxy} \qw}
\newcommand{\multimeasure}[2]{*+<1em,.9em>{\hphantom{#2}} \qw \POS[0,0].[#1,0];p !C *{#2},p \drop\frm<.9em>{-}}
\newcommand{\multimeasureD}[2]{*+<1em,.9em>{\hphantom{#2}} \POS [0,0]="i",[0,0].[#1,0]="e",!C *{#2},"e"+UR-<.8em,0em>;"e"+UL **\dir{-};"e"+DL **\dir{-};"e"+DR+<-.8em,0em> **\dir{-};{"e"+DR+<0em,.8em>\ellipse^{}};"e"+UR+<0em,-.8em> **\dir{-};{"e"+UR-<.8em,0em>\ellipse^{}},"i" \qw}
\newcommand{\control}{*!<0em,.025em>-=-<.2em>{\bullet}}
\newcommand{\controlo}{*+<.01em>{\xy -<.095em>*\xycircle<.19em>{} \endxy}}
\newcommand{\ctrl}[1]{\control \qwx[#1] \qw}
\newcommand{\ctrlo}[1]{\controlo \qwx[#1] \qw}
\newcommand{\targ}{*+<.02em,.02em>{\xy ="i","i"-<.39em,0em>;"i"+<.39em,0em> **\dir{-}, "i"-<0em,.39em>;"i"+<0em,.39em> **\dir{-},"i"*\xycircle<.4em>{} \endxy} \qw}
\newcommand{\qswap}{*=<0em>{\times} \qw}
\newcommand{\multigate}[2]{*+<1em,.9em>{\hphantom{#2}} \POS [0,0]="i",[0,0].[#1,0]="e",!C *{#2},"e"+UR;"e"+UL **\dir{-};"e"+DL **\dir{-};"e"+DR **\dir{-};"e"+UR **\dir{-},"i" \qw}
\newcommand{\ghost}[1]{*+<1em,.9em>{\hphantom{#1}} \qw}
\newcommand{\push}[1]{*{#1}}
\newcommand{\gategroup}[6]{\POS"#1,#2"."#3,#2"."#1,#4"."#3,#4"!C*+<#5>\frm{#6}}

\newcommand{\rstick}[1]{*!L!<-.5em,0em>=<0em>{#1}}
\newcommand{\lstick}[1]{*!R!<.5em,0em>=<0em>{#1}}
\newcommand{\ustick}[1]{*!D!<0em,-.5em>=<0em>{#1}}
\newcommand{\dstick}[1]{*!U!<0em,.5em>=<0em>{#1}}
\newcommand{\Qcircuit}{\xymatrix @*=<0em>}
\newcommand{\link}[2]{\ar @{-} [#1,#2]}
\newcommand{\pureghost}[1]{*+<1em,.9em>{\hphantom{#1}}}


\title{\textbf{\large All Mutually Unbiased Product Bases in Dimension Six}}

\author{Daniel McNulty and Stefan Weigert\\
 Department of Mathematics, University of York\\
York YO10 5DD, UK\\ \\
\small{\tt{dm575@york.ac.uk, stefan.weigert@york.ac.uk}}}

\date{24 March 2012}
\maketitle
\begin{abstract}
All mutually unbiased bases in dimension six consisting of product states only are constructed. Several continuous families of pairs and two triples of mutually unbiased product bases are found to exist but no quadruple. The exhaustive classification leads to a proof that a complete set of seven mutually unbiased bases, if it exists, cannot contain a triple of mutually unbiased product bases. 
\end{abstract}

\section{Introduction}

Mutually unbiased (MU) bases \cite{ivanovic81,wootters+89,Durt+10}
have attracted interest in recent years because their properties seem
to depend dramatically on the dimension $d$ of the quantum system
in hand. If the dimension is given by a prime number $p$, the
state space $\mathbb{C}^{p}$ is known to accommodate a \emph{complete}
set of $(p+1)$ MU bases. Each of these bases consists of $p$ orthonormal states
$\ket{j_a}\in\mathbb{C}^{p}$,
with constant overlap of $1/p$ across different bases, \begin{equation}
| \bk{j_a}{k_b}|^{2}=\frac{1}{p}(1-\delta_{ab})+\delta_{jk}\delta_{ab},\quad
j,k=0\ldots p-1,\,a,b=0\ldots p\,.\label{eq: overlaps}\end{equation}
Complete sets of MU bases also exist for quantum systems with dimension
$d=p^{n}$, where $n$ is a positive integer.
However, for ``composite'' dimensions such as $d\equiv
d_{1}d_{2}\in\{6,10,12,\ldots\}$ complete sets of MU bases seem to be absent. In spite of considerable numerical searches \cite{butterley+07,brierley+08}, computer-algebraic efforts \cite{grassl+04,brierley+09}, and numerical calculations with rigorous error bounds, only three MU bases have been found in dimension six, four less than the maximally allowed number \cite{Jaming+2009}. Thus, the six-dimensional state space of a qubit-qutrit system appears to differ structurally from the state space of a pair of qubits ($d=4)$ or a pair of qutrits $(d=9)$.

One of the few known results in dimension $d=6$ is the impossibility
to extend, by more that one further MU basis, the pair of MU bases consisting
of the standard basis and its dual, the Fourier basis \cite{grassl+04}.
Thus, triples of MU bases are the largest sets to be found in this way.  Another, more recent result \cite{Jaming+2009} states that the Fourier family of Hadamard matrices together with the identity cannot be extended to a MU quadruple. These initial pairs, after non-local equivalence transformations, consist of \emph{product} states only, a fact which has received little attention.

Upon reflection, it seems worthwhile to systematically study MU bases in composite dimensions which contain only product states. In the present paper we carry out a comprehensive study of MU product bases in dimensions six, complementing studies devoted to the entanglement structure of \emph{complete} sets of MU bases \cite{romero+05,Wiesniak+11,Lawrence11}.

More specifically, we will derive an exhaustive list of MU product
bases in dimension six. The restriction to product states goes hand
in hand with local equivalence transformations,  or LETs, consisting of
\emph{local} (anti-) unitary transformations. We will find that in the space $\mathbb{C}^2\otimes\mathbb{C}^3$, there is a considerable number of inequivalent product bases, a limited set of families of MU product \emph{pairs} and just two \emph{triples} of MU product bases. No larger MU product constellations exist. This result  effectively limits the number of MU product bases contained in a hypothetical complete set of MU bases in dimension six. 

The argument will unfold as follows. In Sec. 2 we introduce MU product
bases, specify all local (anti-) unitary transformations which map
a given set of MU product states to an equivalent one, and summarise
relevant properties of MU bases in dimensions two and three. Then,
in Sec. 3, we derive all inequivalent product bases in $\mathbb{C}^{4}$
and $\mathbb{C}^{6}$. Sec. 4 has two results on product vectors required
to be MU to certain given sets of MU product vectors. These results
will be important tools to enumerate all pairs and triples of MU bases in dimension four (Sec. 5) and dimension six (Sec. 6). This classification allows us to conclude, as shown in Sec. 7, that no MU product triple can be part of a complete set of seven MU bases in $d=6$. The final section summarises our findings. 

Readers mainly interested in the results relevant to dimension six are advised to immediately proceed to Sec. 6 after having familiarized themselves with the concept of mutually unbiased product bases presented in Sec. 2. 

\section{MU product bases}

From now on, we will consider quantum systems consisting of two subsystems
with prime dimensions $p$ and $q$, where $p\leq q$. The state space
of such a bipartite system is given by the Hilbert space
$\mathbb{C}^{p}\otimes\mathbb{C}^{q}$
of dimension $d\equiv pq$. Since $q$ is a prime number, there is
a complete set of MU bases of $\mathbb{C}^{q}$, and we will denote its
$q(q+1)$ states by\begin{equation}
\begin{array}{cc}
\ket{J_b}\in\mathbb{C}^{q}\,, & J=0\ldots q-1,\,b=0\ldots
q\,.\end{array}\label{eq: defineMUStatesCq}\end{equation}
The $q$ states $\{\ket{J_{b}}\}$ form one orthonormal basis
labelled by $b$, and states taken from two distinct bases are MU,
in analogy to Eq. (\ref{eq: overlaps}). Given complete sets of MU bases in
$\mathbb{C}^{p}$ and $\mathbb{C}^{q}$, respectively, we now construct $(p+1)$
MU\emph{ product} bases of the space $\mathbb{C}^{p}\otimes\mathbb{C}^{q}$. To
do so, we pair each MU basis of the space $\mathbb{C}^{p}$ with a (different)
basis of $\mathbb{C}^{q}$ and, within each pair, we tensor each state
of the first basis with a (different) state of the second one. This
procedure results in $pq(p+1)$ \emph{product} states \begin{equation}
\ket{j_{a}}\otimes\ket{J_{a}}\equiv\ket{j_{a},J_{a}}\, ,
\label{eq:product bases}\end{equation}
forming $(p+1)$ MU bases $\{\ket{j_{a},J_{a}}, a=0\ldots p\}$
of the space $\mathbb{C}^{p}\otimes\mathbb{C}^{q}$. This is evident upon
calculating the overlaps
\begin{equation}
\left\vert \bk{j_{a},J_{a}}{k_{b},K_{b}}\right\vert ^{2}=\left\vert
\bk{j_{a}}{k_{b}}\right\vert ^{2}\left\vert \bk{J_{a}}{K_{b}}\right\vert
^{2}=\left\{ \begin{array}{ll}
\delta_{jk}\delta_{JK} & \quad\mbox{if \ensuremath{a=b}}\,,\\
\frac{1}{pq} & \quad\mbox{if \ensuremath{a\neq
b}}\,,\end{array}\right.\label{MUBconditionsforproducts}\end{equation}
which are the conditions for bases to be MU in a space of dimension
$pq$.

One can construct MU product bases of the type given in Eq. $(\ref{eq:product bases})$ using Heisenberg-Weyl (HW) operators. In dimension $p$, with $p$ prime, the HW cyclic shift (modulo $p$) and phase operators $X_p$ and $Z_p$, respectively, are defined as
\begin{equation}
\label{HWoperators}
X_p\ket{j}=\ket{j+1} \qquad \mbox{and} \qquad Z_p\ket{j}=\omega^j\ket{j},
\end{equation}
where $\omega=e^{2\pi i/p}$ is a $p^{\text{th}}$ root of unity and $\{\ket{j}\}$ is the standard basis with $j=0\ldots p-1$. Since $p$ is prime, one can construct a complete set of $(p+1)$ MU bases from the eigenbases of the operators $X_p(Z_p)^\ell$ for $0\leq\ell \leq p-1$ \cite{bandyo+02}.

For the composite dimension $d=pq$, we can build a set of $(p+1)$ MU product bases of the Hilbert space $\mathbb{C}^p\otimes\mathbb{C}^q$ with the operators $X_p$ and $Z_p$ acting on the space $\mathbb{C}^p$, and $X_q$ and $Z_q$ on the space $\mathbb{C}^q$. For example, the eigenbases of the operators $X_p\otimes X_q$ and $Z_p\otimes Z_q$ form two MU product bases, which we call a Heisenberg-Weyl pair. One can also construct HW bases with the operators $X_{pq}$ and $Z_{pq}$ on the space $\mathbb{C}^{pq}$, however, these do not necessarily form product bases. Since we are concerned with product bases in this paper, we define the HW operators on the space $\mathbb{C}^p\otimes \mathbb{C}^q$ such that their eigenstates are product states. Note that we do not limit the construction of MU bases to the eigenbases of HW operators, i.e. $\{\ket{j_a,J_a}\}$ in (\ref{eq:product bases}) can be \emph{any} product basis.

Each basis $\{\ket{j_{a},J_{a}}\}$ is a \emph{direct} product basis
of the space $\mathbb{C}^{p}\otimes\mathbb{C}^{q}$ since \emph{each} state
$\ket{j_{a}},j=0\ldots p-1$, of the $a^{\text{th}}$ basis in $\mathbb{C}^{p}$
is multiplied with \emph{every} state $\ket{J_{a}},J=0\ldots q-1$, of the
$a^{\text{th}}$ basis of $\mathbb{C}^{q}$. Direct product bases
are, however, only a subset among all product bases: \emph{indirect}
product bases \cite{Wiesniak+11} result if the states being tensored
stem from more than one basis of the space $\mathbb{C}^{p}$ (or
$\mathbb{C}^{q}$).
The four states\begin{equation}
\left\{
\ket{0_{z},0_{z}},\ket{0_{z},1_{z}},\ket{1_{z},0_{x}},\ket{1_{z},1_{x}}\right\}
\label{eq:indirectproductbasisexample}\end{equation}
provide a simple example of an \emph{indirect} product basis in dimension
four since two different bases of the second space, $\{\ket{j_{z}}\}$
and $\{\ket{j_{x}}\}$, occur in the construction. The matrix representation
of a \emph{direct} product basis in dimension $d=pq$ is given by
the tensor product of two matrices, each representing a basis of the
spaces $\mathbb{C}^{p}$ and $\mathbb{C}^{q}$, respectively. The
matrix representation of an \emph{indirect} product basis cannot be
written as a tensor product of two matrices.

Conceptually, the distinction between direct and indirect product
bases is not linked to MU bases: instead of using $\{\ket{j_{z}}\}$
and $\{\ket{j_{x}}\}$ in (\ref{eq:indirectproductbasisexample})
any other pair of bases of $\mathbb{C}^{2}$ would also define an
indirect product basis. Indirect product bases are important since
they have been found to exhibit a degree of non-locality in the absence
of entanglement \cite{bennet+1999}.

In this paper, we will be concerned exclusively with product bases
of the spaces $\mathbb{C}^{2}\otimes\mathbb{C}^{2}$ and
$\mathbb{C}^{2}\otimes\mathbb{C}^{3}$.
To simplify the construction of all different MU product bases, we
will now introduce equivalence relations which respect the structure
of product states, followed by a brief reminder of the properties
of MU bases in $\mathbb{C}^{2}$ and $\mathbb{C}^{3}$ following conventions used in \cite{brierley+10}.

\subsection{Local equivalence transformations
\label{sub:Local-equivalence-transformations}}

Given a set of MU bases on the space $\mathbb{C}^{p}$, we obtain
another set by applying one single unitary transformation to all states
simultaneously. The scalar products between the states of the MU bases
do not change under this transformation so that we deal indeed with
a second set of MU bases, factually \emph{different} from the initial
set but \emph{equivalent} to it. By not distinguishing between equivalent
MU bases, their enumeration is greatly simplified. When representing MU
bases by Hadamard matrices, the concept of a standard (or dephased) form
emerges naturally (see \cite{bengtsson+07}, for example). To enumerate all MU
\emph{product} bases it will
be helpful not to distinguish those sets of MU product bases which
can be transformed into each other by \emph{local} equivalence transformations,
or LETs, for short. LETs are defined by the requirement that they
preserve the product structure of all states. If there is no LET transforming
two given sets of MU product states into each other they will be called
\emph{locally} \emph{inequivalent}, or just \emph{inequivalent}. It may still be
possible to transform them into each other by \emph{non-local} transformations.

We now list all LETs for a bipartite quantum system with Hilbert space
$\mathbb{C}^{p}\otimes\mathbb{C}^{q}$. Suppose we are given sets
of $(r+1)$ MU bases
$\{\mathcal{B}_{0},\mathcal{B}_{1},\ldots,\mathcal{B}_{r}\}$
that contain only product states. Explicitly, the $\rho^{\text{th}}$
basis, with $\rho=0\ldots r$, consists of $d=pq$ product states 
$\ket{n_\rho,N_\rho}$, $n\equiv N \in \{ 1, 2, \ldots , d\}$,
where $\ket{n_{\rho}}\in\mathbb{C}^{p}$ and 
$\ket{N_{\rho}}\in\mathbb{C}^{q}$.
Any combination of the following five operations maps the given set of MU bases into a locally\emph{ }equivalent set:
\begin{enumerate}
\item a \emph{local unitary} transformation $\hat{u}\otimes\hat{U}$ effecting
\begin{equation}
\mathcal{B}_{\rho}\rightarrow\mathcal{B}_{\rho}^{\prime}
= \hat{u}\otimes\hat{U} \mathcal{B}_{\rho}
\equiv
\Bigl\{\ldots,\ket{\hat{u}n_{\rho}}\otimes\ket{\hat{U}N_{\rho}},\ldots\Bigr\}\,,\label{local
unitary}\end{equation}
 which leaves invariant the value of all scalar products;
\item the multiplication of all states within a basis by possibly different \emph{phase factors} such that \begin{equation}
\mathcal{B}_{\rho}\rightarrow\mathcal{B}_{\rho}^{\prime}=\Bigl\{\ldots,e^{i\phi_{n}^{\rho}}\ket{n_{\rho},N_{\rho}},\ldots\Bigr\}\,;\label{phasefactor}\end{equation}
these transformations exploit the fact that the overall phase of a
quantum state has no physical significance and automatically drops
out from the conditions defining MU bases. It is worth noting that a single
phase factor $e^{i\phi}$ can dephase \emph{both} states of a product: let
$\phi\equiv\phi^{\prime}+\phi^{\prime\prime}$ to find
$e^{i\phi}\ket{n_\rho,N_\rho}=(e^{i\phi^{\prime}}\ket{n_\rho})\otimes(e^{i\phi^{\prime\prime}}\ket{N_\rho})$;
\item \emph{permutations} of the product states within each basis; as an example, consider the permutation of states
$\ket{n_{\rho},N_{\rho}}$
and $\ket{n^{\prime}_{\rho},N^{\prime}_{\rho}}$ in
the $\rho^{\text{th}}$ basis \label{permutation} \begin{equation}
\Bigl\{\ldots,\ket{n_{\rho},N_{\rho}},\ldots,\ket{n^{\prime}_{\rho},N^{\prime}_{\rho}},\ldots\Bigr\}\longrightarrow\Bigl\{\ldots,\ket{n^{\prime}_{\rho},N^{\prime}_{\rho}},\ldots,\ket{n_{\rho},N_{\rho}},\ldots\Bigr\}\,,\end{equation}
 which amounts to relabelling the elements within each basis;
\item the \emph{local complex conjugations} $\hat{k}\otimes\hat{I}$ and
$\hat{I}\otimes\hat{K}$ (anti-unitary operations defined with respect
to the standard bases in $\mathbb{C}^{p}$ and $\mathbb{C}^{q}$,
respectively), and thus their product $\hat{k}\otimes\hat{K}$; for
example, applying $\hat{k}\otimes\hat{I}$ \label{complex conjugation}
\begin{equation}
\mathcal{B}_{\rho}\rightarrow\mathcal{B}_{\rho}^{\prime}=\Bigl\{\ldots,\ket{n^*_{\rho},N_{\rho}},\ldots\Bigr\}\,,\label{complex
conjugation1}\end{equation}
swaps all scalar products resulting from the first factors without
changing their numerical values;
\item \emph{pairwise exchanges} of two bases, which amounts to relabelling
the bases.
\end{enumerate}
We now briefly discuss some important properties of LETs. First, they
represent a true subset of all equivalence transformations in a space of dimension $pq$: no LET maps an indirect product basis to a direct one while a general unitary equivalence transformation can send any orthonormal basis to any other. Second, we will find indirect product bases which cannot be transformed into each other
by LETs, i.e. \emph{locally inequivalent} product bases. As a result, the idea of a \emph{unique} standard or dephased
form of MU bases is less straightforward for MU product bases. We
define a standard form in the following way: the first basis $\mathcal{B}_{0}$,
be it direct or indirect, contains the states $\{\ket{j_{z}}\}$
of the space $\mathbb{C}^{p}$ and the states $\{\ket{J_{z}}\}$ of
the space $\mathbb{C}^{q}$; the second basis $\mathcal{B}_{1}$ contains
the state $\ket{0_{x},0_{x}}$, and all other states in the remaining
bases are dephased using the transformation defined in $(\ref{phasefactor})$.
Superficially, LETs remind one of local operations with classical
communication, or LOCCs \cite{nielsen99}. However, the presence of
anti-unitary operations
rather suggests a link with Wigner's theorem about symmetry transformations
leaving transition probabilities invariant \cite{wigner31}, for the special
case of a universe populated with product states only. Finally, it is
straightforward to generalise LETs to
$n$-partite systems residing in product states only.

It is often convenient to represent an MU product basis in $\mathbb{C}^{pq}$ as a complex Hadamard matrix of dimension $(pq\times pq)$, with each product state corresponding to one column. The bases $\{\mathcal{B}_{0},\mathcal{B}_{1},\ldots,\mathcal{B}_{r}\}$
then turn into a set of $(r+1)$ matrices, on which the five transformations
above act in the following way. The first LET is a \emph{local unitary},
given by the Kronecker product of two unitary matrices, applied to
all matrices from the left; the second LET corresponds to \emph{diagonal unitary} transformations acting from the right; \emph{unitary permutation matrices} acting from the right implement
the third type of LET, while the effect of the \emph{local complex
conjugations}
must be worked out by writing down each product state individually.

\subsection{MU bases in dimensions two and three
\label{sub:All-MU-basesdim2and3}}

Given a pair of MU bases in the vector space $\mathbb{C}^{2}$, we
can always map the first basis to the standard basis $\{\ket{j_{z}}\}$
by a suitable unitary transformation $\hat{u}\in SU(2).$ Being MU
to the first basis, the states of the second basis now must have the
form\begin{equation}
\ket
a=\frac{1}{\sqrt{2}}(\ket{0_{z}}+e^{i\lambda}\ket{1_{z}})\equiv\hat{r}_{\lambda}\ket
+\:,\quad\ket{a^{\dagger}}=\hat{r}_{\lambda}\ket
-\:,\label{eq:C2MUvectors}\end{equation}
where $\{\ket{\pm}\}\equiv\{\ket{j_{x}}\}$ is the $x$-eigenbasis,
and the operator $\hat{r}_{\lambda},\lambda\in[0,\pi)$, represents
a rotation by an angle $\lambda$ about the $z$-axis. Since any such
rotation leaves the standard basis $\{\ket{j_{z}}\}$ unchanged, the
second MU basis can be transformed into $\{\ket{j_{x}}\}$. The matrix
representation of the resulting pair of MU bases reads\begin{equation}
\left\{ I;F_{2}\right\} \equiv\left\{ \left(\begin{array}{rr}
1 & 0\\
0 & 1\end{array}\right);\frac{1}{\sqrt{2}}\left(\begin{array}{rr}
1 & 1\\
1 & -1\end{array}\right)\right\} \,.\end{equation}
All other pairs of MU bases of the space $\mathbb{C}^{2}$ are, in
fact, equivalent to this one. A third basis MU to these two bases
consists of the states given in Eq. (\ref{eq:C2MUvectors}) if
$\lambda=\pm\pi/2$,
producing $\{\ket{j_{y}}\}$. Thus, all pairs of MU bases in $\mathbb{C}^{2}$
are equivalent to $\{\ket{j_{z}};\ket{j_{x}}\}$, and all triples
are equivalent to $\{\ket{j_{z}};\ket{j_{x}};\ket{j_{y}}\}$, as is
well known.

In dimension three, one of two given MU bases can always be mapped
to the standard basis $\{\ket{J_{z}},J=0,1,2\}$, so that the second basis consists of states of the form\begin{equation}
\ket
A=\frac{1}{\sqrt{3}}(\ket{0_{z}}+e^{i\xi}\ket{1_{z}}+e^{i\eta}\ket{2_{z}})\,,\quad\xi,\eta\in[0,2\pi)\,,\label{eq:C3MUvectorA}\end{equation}
exploiting the fact that the overall phase of a quantum state has
no physical meaning. One can construct three states of this form which
are pairwise orthogonal: writing \begin{equation}
\ket{A^{\perp}}=\frac{1}{\sqrt{3}}(\ket{0_{z}}+\gamma
e^{i\xi}\ket{1_{z}}+\delta
e^{i\eta}\ket{2_{z}})\,,\quad|\gamma|=|\delta|=1\,,\end{equation}
the condition $\bk A{A^{\perp}}=0$ implies $\gamma+\delta=-1$. A
geometric argument in the complex plane implies either $\gamma=\omega$
and $\delta=\omega^{2}$, or $\gamma=\omega^{2}$ and $\delta=\omega$,
where $\omega=e^{2\pi i/3}$ is a third root of unity. We denote the resulting
basis by \begin{equation}
\{\ket{A},\ket{A^{\perp}},\ket{A^{\perp\!\!\!\perp}}\}=\{\hat{R}_{\xi,\eta}\ket{J_{x}}\}\,,\label{eq:RJx}\end{equation}
where the triple $\{\ket{J_{x}}\}\equiv\{\ket{J_{x}},J=0,1,2\}$ consists
of the eigenstates of the shift operator $\hat{X}_3$, and the operator
$\hat{R}_{\xi,\eta}$ is diagonal in the $z$-basis such that $\ket
A\equiv\hat{R}_{\xi,\eta}\ket{0_{x}}$,
cf. Eq. (\ref{eq:C3MUvectorA}). The free parameters $\xi,\eta$ in
the pairs of MU bases $\{\ket{J_{z}};\hat{R}_{\xi,\eta}\ket{J_{x}}\}$
can be removed by a suitable redefinition of the phases of the states
in the standard basis $\{\ket{J_{z}}\}$. Thus, all pairs of MU bases
of $\mathbb{C}^{3}$ are equivalent to the pair $\{\ket{J_{z}};\ket{J_{x}}\}$
which may be represented by 
\begin{equation}
\left\{ I;F_3\right\} =\left\{ \left(\begin{array}{ccc}
1 & 0 & 0\\
0 & 1 & 0\\
0 & 0 & 1\end{array}\right);\frac{1}{\sqrt{3}}\left(\begin{array}{ccc}
1 & 1 & 1\\
1 & \omega & \omega^{2}\\
1 & \omega^{2} & \omega\end{array}\right)\right\} \,,\label{zandxmatrix}\end{equation}
where $F_3\equiv H_x$ is the Fourier matrix in $\mathbb{C}^{3}$. 
Note that two more orthonormal bases of states MU to the pair
$\{\ket{J_{z}};\ket{J_{x}}\}$
emerge if one sets either $e^{i\xi}=e^{i\eta}\equiv\omega$ or
$e^{i\xi}=e^{i\eta}\equiv\omega^{2}$
in Eq. (\ref{eq:RJx}). We will denote these bases by $\{\ket{J_{y}}\}$
and $\{\ket{J_{w}}\}$, respectively, and their matrix representations
are given by \begin{equation}
H_{y}=\frac{1}{\sqrt{3}}\left(\begin{array}{ccc}
1 & 1 & 1\\
\omega & \omega^{2} & 1\\
\omega & 1 & \omega^{2}\end{array}\right)\,,\quad
H_{w}=\frac{1}{\sqrt{3}}\left(\begin{array}{ccc}
1 & 1 & 1\\
\omega^{2} & 1 & \omega\\
\omega^{2} & \omega & 1\end{array}\right)\,,\label{yandwmatrix}\end{equation}
which are also MU with respect to each other. The matrices $H_x,H_y$ and $H_w$ are complex $(3\times3)$ Hadamard matrices, i.e. they are unitary and the moduli of all their entries are equal to $1/\sqrt{3}$.

Two \emph{triples} of MU bases now result from adding either $\{\ket{J_{y}}\}$
or $\{\ket{J_{w}}\}$ to the pair $\{\ket{J_{z}};\ket{J_{x}}\}$.
These triples are equivalent to each other as follows from taking the
complex conjugate (defined in the $z$-basis) of the triple
$\{\ket{J_{z}};\ket{J_{x}};\ket{J_{y}}\}$:
the complex conjugation only affects the ordering of states within $\{\ket{J_{x}}\}$
while $\{\ket{J_{y}}\}$ turns into $\{\ket{J_{w}}\}$. Thus we conclude that the triples are indeed equivalent which we express formally by writing \begin{equation}
\{\ket{J_{z}};\ket{J_{x}};\ket{J_{y}}\}\sim\{\ket{J_{z}};\ket{J_{x}};\ket{J_{w}}\}\,.\end{equation}
Consequently, all MU triples are equivalent to the triple
$\{\ket{J_{z}};\ket{J_{x}};\ket{J_{y}}\}$,
and the complete set of four MU bases in $\mathbb{C}^{3}$ is also
unique, as is well known.

\section{Constructing product bases in dimensions four and six}

The first step towards an exhaustive list of pairs and triples of
MU product bases in dimension six is to construct all locally inequivalent
product bases in $\mathbb{C}^{2}\otimes\mathbb{C}^{3}$. Once these
are known, the requirement of any two such bases to be MU will impose
further constraints. It will be helpful to initially carry out this construction in dimension four. Thus, we will first derive all inequivalent product bases of the space $\C^2\otimes\C^2$, followed by a similar construction for a six-dimensional space.

\subsection{All product bases in $d=4$ \label{sub:all-product-bases-d=00003D4}}

We now show that each product basis in $d=4$ is equivalent either
to the standard \emph{direct} product basis or to a member of two families of
\emph{indirect} product bases, each depending on two real parameters. Any
(orthonormal) product basis
in the space $\mathbb{C}^{2}\otimes\mathbb{C}^{2}$ must have the
form \begin{equation}
\Bigl\{\ket{\psi_{1},\phi_{1}},\,\ket{\psi_{2},\phi_{2}},\,\ket{\psi_{3},\phi_{3}},\,\ket{\psi_{4},\phi_{4}}\Bigr\}\,,\end{equation}
where $\ket{\psi_{n}},\ket{\phi_{n}}\in\mathbb{C}^{2}$ for $n=1\ldots4$.
The conditions \begin{equation}
\bk{\psi_{n},\phi_{n}}{\psi_{n^{\prime}},\phi_{n^{\prime}}}=\bk{\psi_{n}}{\psi_{n^{\prime}}}\bk{\phi_{n}}{\phi_{n^{\prime}}}=\delta_{nn^{\prime}}\,,\quad
n,n^{\prime}=1\ldots4\,,\end{equation}
imply that at least two states of the first factor must be orthogonal. However, no three
orthogonal states exist in $\mathbb{C}^{2}$, so that upon calling
$\ket{\psi_{1}}\equiv\ket a$ we must have \begin{equation}
\Bigl\{\ket{a,\phi_{1}},\,\ket{a^{\perp},\phi_{2}},\,\ket{\psi_{3},\phi_{3}},\,\ket{\psi_{4},\phi_{4}}\Bigr\}\,,\label{generalbasisd=4}\end{equation}
with $\ket{\psi_{2}}=\ket{a^{\perp}}$ being the unique state orthogonal
to $\ket a$. Now we need to consider two separate cases: we can have
either $\ket{\psi_{3}}=\ket a$ (or, equivalently,
$\ket{\psi_{3}}=\ket{a^{\perp}}$)
or $\ket{\psi_{3}}=\ket b$ such that $0<|\bk ab|<1$, meaning that
the state $\ket b$ is neither a multiple of the state $\ket a$ nor
orthogonal to it; we call such a vector $\ket b$ \emph{skew} to $\ket a$. 

By a simple argument using the restrictions imposed by the orthogonality conditions, one finds that three different bases result:
\begin{align}
\mathcal{B}_{0}&=\Bigl\{\ket{a,A},\,\ket{a,A^{\perp}},\,\ket{a^{\perp},A},\,\ket{a^{\perp},A^{\perp}}\Bigr\},\label{B0ind=4}\\
\mathcal{B}_{1}&=\Bigl\{\ket{a,A},\,\ket{a,A^{\perp}},\,\ket{a^{\perp},B},\,\ket{a^{\perp},B^{\perp}}\Bigr\}\,,\label{B1ind=4}\\
\mathcal{B}_{2}&=\Bigl\{\ket{a,A},\,\ket{a^{\perp},A},\,\ket{b,A^{\perp}},\,\ket{b^{\perp},A^{\perp}}\Bigr\}\,.\label{B2ind=4}\end{align}
The basis $\mathcal{B}_{0}$ is a direct product basis while the bases $\mathcal{B}_{1}$ and $\mathcal{B}_{2}$ are not. After performing suitable LETs, we can thus summarise the complete list of product bases in dimension four as follows. 
\begin{lem}
\label{thm:Any-orthonormal-productd=00003D4} Any orthonormal product
basis of the space $\mathbb{C}^{2}\otimes\mathbb{C}^{2}$ is equivalent
to a member of one of the families \begin{align}
\mathcal{I}_{0} & =\{\ket{j_{z},k_{z}}\}\,, \nonumber \\
\mathcal{I}_{1} & =\{\ket{0_{z},k_{z}},\ket{1_{z},\hat{u}k_{z}}\}\,,\nonumber \\
\mathcal{I}_{2} & =\{\ket{j_{z},0_{z}},\ket{\hat{v}j_{z},1_{z}}\}\,,\label{eq:
I2dim4}\end{align}
where the operators $\hat{u},\hat{v}\in SU(2)$ act on the space
$\mathbb{C}^{2}$
such that the states $\ket{0_{z}}$ and $\hat{u}\ket{0_{z}}$, as
well as the states $\ket{0_{z}}$ and $\hat{v}\ket{0_{z}}$, are skew.
\end{lem}
Note that the parameters on which the operators depend have been chosen in such a way that no product basis occurs more than once.
A number of LETs (cf Sec. \ref{sub:Local-equivalence-transformations})
have been used to bring the bases into the form given in the lemma.
The basis $\mathcal{B}_{0}$ in (\ref{B0ind=4}) has been
mapped to $\mathcal{I}_{0}$ by means of a transformation
$\hat{u}_{1}\otimes\hat{u}_{2}$
such that $\hat{u}_{1}$ maps the pair of states $\{\ket a,\ket{a^{\perp}}\}$
to the standard basis $\{\ket{j_{z}}\}$ of $\mathbb{C}^{2}$, and
$\hat{u}_{2}$ is defined analogously. Thus, the bases $\mathcal{B}_{0}$
and $\mathcal{I}_{0}$ are equivalent to each other. We apply a similar
transformation to the basis $\mathcal{B}_{1}$ in (\ref{B1ind=4})
mapping two of the bases to the standard basis. The freedom to choose
a third basis, associated with the pair $\{\ket B,\ket{B^{\perp}}\}$,
is represented in $\mathcal{I}_{1}$ by the undetermined unitary operator
$\hat{u}$ acting on the standard basis. The same reasoning brings
$\mathcal{B}_{2}$ into the form (\ref{eq: I2dim4}) except that the
roles of the two spaces are swapped. Since a complex conjugation reflects
points on the Bloch sphere about the $xz$-plane, only half of all
the unitaries $\hat{u}$ (and $\hat{v}$) need to be considered in
Lemma \ref{thm:Any-orthonormal-productd=00003D4}. In other words,
the bases associated with the unitaries $\hat{u}$ and $\hat{u}^{*}$,
given by the complex conjugate of the matrix representing $\hat{u}$
in the $z$-basis, coincide.

The symmetry of the space $\mathbb{C}^{2}\otimes\mathbb{C}^{2}$ is
reflected in the fact that we found two bases $\mathcal{I}_{1}$ and
$\mathcal{I}_{2}$ which are identical except for the order of the
factors. If we stick with the idea that LETs dictate whether two product
bases are equivalent to each other, we need to consider these bases as
inequivalent. Thus, the complete set of product bases
consists of two families each of which depends on two parameters due to the $SU(2)$-transformations $\hat{u}$ and $\hat{v}$. Not all three parameters of a
transformation in $SU(2)$ are relevant since the overall phase of quantum states is physically irrelevant: each pair of opposite points on the Bloch sphere defines an orthonormal basis of $\mathbb{C}^{2}$
so that the set of all bases depends on only two real parameters. Note that the sets $\mathcal{I}_{1}$ and $\mathcal{I}_{2}$ of Lemma \ref{thm:Any-orthonormal-productd=00003D4} are both connected to the product basis $\mathcal{I}_{0}$. 

\begin{figure}[ht]
\newcommand{\dlstick}[1]{*!U!<1.4em,.6em>=<0em>{#1}}
\newcommand{\drstick}[1]{*!U!<-.5em,.6em>=<0em>{#1}}
\centerline{
 \Qcircuit @C=1em @R=0.7em @!R
{\dlstick{\mathcal{I}_{0}} & \ctrl{1} & \qw &\drstick{\mathcal{I}_{1}} \\
 & \gate{\hat{u}} & \qw } \qquad \qquad \qquad
\Qcircuit @C=1em @R=0.7em @!R
{\dlstick{\mathcal{I}_{0}} & \gate{\hat{v}} & \qw & \drstick{\mathcal{I}_{2}} \\
 & \ctrl{-1} & \qw  }
}
\caption{Two quantum circuits to create  the product bases $\mathcal{I}_1$ and
$\mathcal{I}_2$, respectively; the unitaries $\hat{u}$ and $\hat{v}$ only act
on the target qubit if the control qubit is in the state $\ket{1_z}$.}
\label{dim4circuits}
\end{figure}
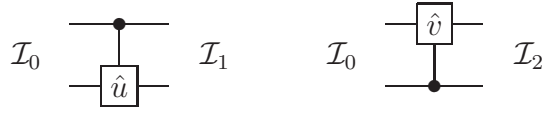

The symmetry becomes particularly obvious if we represent the bases
of Lemma \ref{thm:Any-orthonormal-productd=00003D4} by quantum circuits.
The idea is to visualise the operation needed to map the states of
the standard product basis $\mathcal{I}_{0}$ into the desired product
basis by means of a quantum gate. This is always possible since any
two orthonormal bases are connected by a unitary operation. Obviously,
the trivial gate, described by the identity $\hat{I},$ maps the four
vectors of the standard product basis to itself. Fig. (\ref{dim4circuits})
shows that
(non-local) controlled-$\hat{u}$ and controlled-$\hat{v}$ gates are required
to output the bases $\mathcal{I}_{1}$ and $\mathcal{I}_{2}$, respectively.
As expected, the two circuits are identical upon swapping the qubits.

\subsection{All product bases in $d=6$ \label{sub:All-product-basesdim6}}

To construct all product bases in dimension six we use the same method
as in dimension four. Any product basis in the space
$\mathbb{C}^{2}\otimes\mathbb{C}^{3}$
takes the form \begin{equation}
\Bigl\{\ket{\psi_{1},\Psi_{1}},\,\ket{\psi_{2},\Psi_{2}},\,\ket{\psi_{3},\Psi_{3}},\,\ket{\psi_{4},\Psi_{4}},\,\ket{\psi_{5},\Psi_{5}},\,\ket{\psi_{6},\Psi_{6}}\Bigr\}\,,\label{eq:
candidatebasisindim6}\end{equation}
with states $\ket{\psi_{n}}\in\mathbb{C}^{2}$ and
$\ket{\Psi_{n}}\in\mathbb{C}^{3}$
for $n=1\ldots6\,$, satisfying the orthogonality conditions \begin{equation}
\bk{\psi_{n},\Psi_{n}}{\psi_{n^{\prime}},\Psi_{n^{\prime}}}=\bk{\psi_{n}}{\psi_{n^{\prime}}}\bk{\Psi_{n}}{\Psi_{n^{\prime}}}=\delta_{nn^{\prime}}\,,\quad
n,n^{\prime}=1\ldots6\,.\label{eq:orthoconddim6}\end{equation}
The states $\ket{\psi_{n}}\,,n=1\ldots6$, in (\ref{eq: candidatebasisindim6})
must contain at least \emph{two} (not necessarily different)  pairs of orthogonal states.
If they do not, the orthogonality conditions require \emph{four}
orthogonal states in $\mathbb{C}^{3}$, which do not exist. In fact,
the remaining two states in $\mathbb{C}^{2}$ must also be orthogonal,
which implies that the product bases of $\mathbb{C}^{6}$ will come
in three flavours. The states $\ket{\psi_{n}}\,,n=1\ldots6$, fall
into three pairs of states consisting of either three, two, or only one
pair of orthonormal bases. The following lemma summarises the results
of the detailed arguments given in Appendix \ref{sec:Appendixproductbasesdim6}.
\begin{lem}
\label{thm:product-bases-d=00003D6} Any orthonormal product basis
of the space $\mathbb{C}^{2}\otimes\mathbb{C}^{3}$ is equivalent
to a member of one of the families \begin{align}
\mathcal{I}_{0} & =\{\ket{j_{z},J_{z}}\}\,, \nonumber \\
\mathcal{I}_{1} & =\{\ket{0_{z},J_{z}},\ket{1_{z},\hat{U}J_{z}}\}\,,\nonumber \\
\mathcal{I}_{2} & =\{\ket{j_{z},0_{z}},\ket{\hat{u}0_{z},1_{z}},\ket{\hat{u}0_{z},2_{z}},\ket{\hat{u}1_{z},\hat{V}1_{z}},\ket{\hat{u}1_{z},\hat{V}2_{z}}\}\,, \nonumber 
\\
\mathcal{I}_{3} & =\{\ket{j_{z},0_{z}},\ket{\hat{v}j_{z},1_{z}},\ket{\hat{w}j_{z},2_{z}}\}\,, \label{eq:I3}
\end{align}
with $j=0,1$ and $J=0,1,2$; the operators $\hat{u},\hat{v},\hat{w}\in
SU(2)$
and $\hat{U},\hat{V}\in SU(3)$ act on $\mathbb{C}^{2}$ and $\mathbb{C}^{3}$,
respectively, with $\hat{V}$ leaving the the state $\ket{0_{z}}$
invariant; the parameters of the operators $\hat{u},\ldots,\hat{V}$
are chosen in such a way that no product basis occurs more than once.
\end{lem}
Without any restrictions on the five unitary operators $\hat{u},\ldots,\hat{V}$
some product
bases would occur more than once in this list. For example, if
$\hat{U}\equiv\hat{I}$,
the basis $\mathcal{I}_{1}$ turns into $\mathcal{I}_{0}$; similarly,
the bases associated with $\hat{U}$ and $\hat{U}^{*}$ are identical.
We could remove such multiple occurrences by appropriately restricting
the unitary operators but it is rather cumbersome to do so and not
particularly informative.

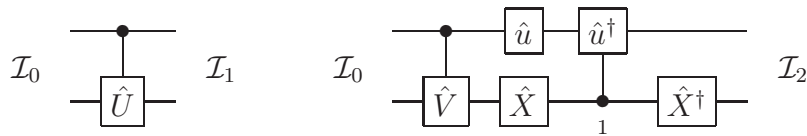
\begin{figure}[ht]
\newcommand{\dlstick}[1]{*!U!<1.4em,.8em>=<0em>{#1}}
\newcommand{\drstick}[1]{*!U!<-.5em,.8em>=<0em>{#1}}
\centerline{
\Qcircuit @C=1em @R=0.7em @!R { \dlstick{\mathcal{I}_{0}} & \ctrl{1} & \qw &
\drstick{\mathcal{I}_{1}}\\  & \gate{\hat{U}} & \qw }\qquad \qquad \qquad
\Qcircuit @C=1em @R=0.7em @!R {
\dlstick{\mathcal{I}_{0}} & \ctrl{1} & \gate{\hat{u}^{\vphantom\dagger}} & \gate{\hat{u}^\dagger} & \qw & \qw & \drstick{\mathcal{I}_{2}} \\ 
& \gate{\hat{V}^{\vphantom{\dagger}}} & \gate{\hat{X}^{\vphantom{\dagger}}} & \ctrl{-1} & \gate{\hat{X}^\dagger} & *!U!<4.64em,.6em>=<0em>{\text{\scriptsize1}}\qw }}
\caption{Quantum circuits for a qubit (upper wire) and qutrit (lower wire) to create the bases $\mathcal{I}_1$ and $\mathcal{I}_2$, respectively; the
controlled-$\hat{U}$ and controlled-$\hat{V}$ gate act on the qutrit only if the control qubit is in the state $\ket{1_z}$; the unitary $\hat{u}^\dagger$, the adjoint of $\hat{u}$, acts on the qubit only when the control qutrit is in the state $\ket{1_z}$; and the operator $\hat{X}$ acts as a shift on the standard basis of $\mathbb{C}^3$. }
\label{dim6bases1and2circuits}
\end{figure}

\begin{figure}[ht]
\newcommand{\dlstick}[1]{*!U!<1.4em,.8em>=<0em>{#1}}
\newcommand{\drstick}[1]{*!U!<-.5em,.8em>=<0em>{#1}}
\centerline{
\Qcircuit @C=1em @R=0.7em @!R {
\dlstick{\mathcal{I}_{0}} & \gate{\hat{w}} & \qw & \gate{\hat{v}} &
\qw & \drstick{\mathcal{I}_{3}} \\
& \ctrl{-1}*!U!<-1.69em,.6em>=<0em>{\text{\scriptsize1}}&\gate{\hat{X}} & \ctrl{-1}  &
*!U!<1.55em,.6em>=<0em>{\text{\scriptsize1}} \qw
} }
\caption{A quantum circuit for a qubit (upper wire) and qutrit (lower wire)  to create the basis $\mathcal{I}_3$; the unitaries $\hat{v}$ and $\hat{w}$ act on the qubit only if the control qutrit is in the state $\ket{1_z}$, and the operator $\hat{X}$ acts as a shift on the standard basis of $\mathbb{C}^3$.}
\label{dim6basis3circuit}\end{figure}
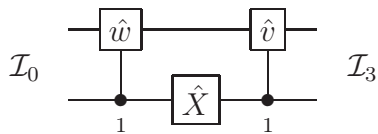

Compared to dimension four, the number of families of indirect product
bases have increased, and they contain transformations
generated by elements of the group $SU(3)$. Clearly, there is no scope
for symmetry under exchanging the two spaces of the product
$\mathbb{C}^{2}\otimes\mathbb{C}^{3}$.
The families $\mathcal{I}_{1}$ to $\mathcal{I}_{3}$ each depend on a number of free parameters: $\mathcal{I}_{1}$ has six free parameters due to the unitary $\hat{U}$; two free parameters are associated with each $SU(2)$-transformation present in $\mathcal{I}_{3}$, while $\mathcal{I}_{2}$ is a five-parameter family -- the transformations due to $\hat{V}$, which is effectively an $SU(2)$-transformation, brings not only two but \emph{three} parameters because the overall phase of the states in the two-dimensional subspace spanned by $\ket{1_{z}}$ and $\ket{2_{z}}$
does \emph{not} drop out. Figs. (\ref{dim6bases1and2circuits}) and
(\ref{dim6basis3circuit}) show quantum circuits to generate the inequivalent
product bases in dimension six.

\section{Adding MU product states to sets of orthogonal product vectors}

In this section we derive a theorem which will play a crucial role
in the construction of \emph{all} pairs and triples of MU product
bases in dimension four and six. This theorem is inspired by a constraint
on two \emph{direct} product bases to be MU, obtained in
\cite{Wiesniak+11}:
\begin{lem*}
\label{thm:Zeilinger}Two [direct] product bases
$\{\ket{j_{a},J_{a}}\}$ and $\{\ket{k_{b},K_{b}}\}$ in dimension
$d=pq$ are MU if and only if $\ket{j_{a}}$ is MU to $\ket{k_{b}}$
in dimension $p$ and $\ket{J_{a}}$ is MU to $\ket{K_{b}}$ in dimension
$q$.
\end{lem*}
This result does not cover \emph{indirect} bases. To (partly) remedy
this shortcoming, we will present two different ways to generalise
this Lemma. Firstly, we find a constraint on each product vector if
it is to be MU to a specific set of product vectors; this result is
obtained for spaces of arbitrary composite dimension $d=pq$. Secondly,
we derive constraints on a product vector required to be MU to \emph{any}
(direct or indirect) given product basis of the spaces $\mathbb{C}^{4}$
or $\mathbb{C}^{6}$.

Consider $p$ product states $\{\ket{\psi_{i},\Psi},i=1\ldots p\}$
with an orthonormal basis $\{\ket{\psi_{i}},i=1\ldots p\}$ of the
space $\mathbb{C}^{p}$, and with $\ket{\Psi}\in\mathbb{C}^{q}$.
After swapping the two factors in Eq. (\ref{eq:indirectproductbasisexample}),
the product basis $\{\ket{j_{z},0_{z}},\ket{j_{x},1_{z}}\}$, for
example, is seen to consist of two sets of this form. We find that only
particular product states can be MU to such sets of product states.
\begin{lem}
\label{thm:-product-states-MU-to-sets}The product state $\ket{\phi,\Phi}$
in dimension $d=pq$ is MU to the set of orthogonal product states
$\{\ket{\psi_{i},\Psi},i=1\ldots p\}$ if and only if  $\ket{\phi}$
is MU to $\ket{\psi_{i}}\in\mathbb{C}^{p}$ and $\ket{\Phi}$ is MU
to $\ket{\Psi}\in\mathbb{C}^{q}$.
\end{lem}
If $|\bk{\psi_{i}}{\phi}|^{2}=1/p$ and $|\bk{\Psi}{\Phi}|^{2}=1/q$,
then the product states are indeed MU in the space $\mathbb{C}^{pq}$
since it follows that
$|\bk{\psi_{i},\Psi}{\phi,\Phi}|^{2}=|\bk{\psi_{i}}{\phi}|^{2}|\bk{\Psi}{\Phi}|^{2}=1/pq$.
To prove the converse, we assume the product states are MU,
$|\bk{\psi_{i},\Psi}{\phi,\Phi}|^{2}=1/pq$.
Summing over $i=1\ldots p$, we obtain $|\bk{\Psi}{\Phi}|^{2}=1/q$ upon using
the completeness relation $\sum_{i}|\bk{\psi_{i}}{\phi}|^{2}=1$.
This result immediately implies that $|\bk{\psi_{i}}{\phi}|^{2}=1/p\,,i=1\ldots
p$,
also holds.

Note that one can swap the roles of the factors in the tensor product.
Then Lemma \ref{thm:-product-states-MU-to-sets} restricts the form of any product state which is MU
to a set of $q$ orthogonal states $\{\ket{\psi,\Psi_{i}},i=1\ldots q\}$
with an orthonormal basis $\{\ket{\Psi_{i}},i=1\ldots q\}$ of the
space $\mathbb{C}^{q}$, and with $\ket{\psi}\in\mathbb{C}^{p}$.

This result covers the Lemma given at the beginning of this section.
To see this, group the basis $\{\ket{j_{a},J_{a}}\}$ into $q$ sets
of $p$ orthonormal vectors
$\{\ket{j_{a},1_{a}}\}$, $\{\ket{j_{a},2_{a}}\}\ldots$ $\{\ket{j_{a},q_{a}}\}$;
then, by Lemma \ref{thm:-product-states-MU-to-sets}, any product
state $\ket{\phi,\Phi}$ is mutually unbiased to each set of vectors
if and only if the state $\ket{\phi}$ is MU to all states $\ket{j_{a}}$,
and the state $\ket{\Phi}$ is MU to all states $\ket{J_{a}}$. By
replacing the state $\ket{\phi,\Phi}$ with a vector from the basis
$\{\ket{k_{a},K_{a}}\}$ and repeating the argument for all states
in this basis, one arrives at the Lemma for \emph{direct} product
bases.

The following generalisation uses the fact that we know all direct
and indirect product bases in dimensions four and six.
\begin{thm}
\label{thm:product-states-MU-to-bases-d=00003D4,6} The product state
$\ket{\phi,\Phi}\in\mathbb{C}^{d},d\equiv pq\leq6,$ is MU to the
product basis $\{\ket{\psi_{i},\Psi_{i}}\}$ with $i=1\ldots pq$,
if and only if $\ket{\phi}$ is MU to $\ket{\psi_{i}}\in\mathbb{C}^{p}$
and $\ket{\Phi}$ is MU to $\ket{\Psi_{i}}\in\mathbb{C}^{q}$.
\end{thm}
We prove this statement by considering the cases $d=4$ and $d=6$
separately:

$\bullet\, d=4$: All product bases in dimension four are given by
the bases $\mathcal{I}_{0},\mathcal{I}_{1}$ and $\mathcal{I}_{2}$,
collected in Lemma \ref{thm:Any-orthonormal-productd=00003D4}. Each
of these bases can be divided into groups of states of the form
$\{\ket{\psi_{j},\Psi},j=1,2\}$,
or $\{\ket{\psi,\Psi_{j}},j=1,2\}$. Thus, Theorem
\ref{thm:product-states-MU-to-bases-d=00003D4,6}
follows immediately from Lemma \ref{thm:-product-states-MU-to-sets}.

$\bullet\, d=6$: It is sufficient to consider the four families of
bases given in Lemma \ref{thm:product-bases-d=00003D6}. Each of the
bases $\mathcal{I}_{0}$, $\mathcal{I}_{1}$ and $\mathcal{I}_{3}$
can be split into sets of the form required to apply Lemma
\ref{thm:-product-states-MU-to-sets};
thus, Theorem \ref{thm:product-states-MU-to-bases-d=00003D4,6} holds
for these bases. To complete the proof, we need to consider the basis
$\mathcal{I}_{2}$ which has no such decomposition. To begin, suppose
that the basis $\mathcal{I}_{2}$ \emph{is} MU to the state $\ket{\phi,\Phi}$.
According to Lemma \ref{thm:-product-states-MU-to-sets} this state
is MU to the pair $\{\ket{j_{z},0_{z}}\}$ if both
$|\bk{\phi}{0_{z}}|^{2}=|\bk{\phi}{1_{z}}|^{2}=1/2$
and $|\bk{\Phi}{0_{z}}|^{2}=1/3$ hold. The state $\ket{\phi,\Phi}$
also needs to satisfy \begin{equation}
|\bk{\phi}{\hat{u}0_{z}}|^{2}|\bk{\Phi}{1_{z}}|^{2}=|\bk{\phi}{\hat{u}0_{z}}|^{2}|\bk{\Phi}{2_{z}}|^{2}=\frac{1}{6}\,;\label{eq:
|MUsubspace2}\end{equation}
adding these two constraints we find \begin{equation}
|\bk{\phi}{\hat{u}0_{z}}|^{2}\Bigl(|\bk{\Phi}{1_{z}}|^{2}+|\bk{\Phi}{2_{z}}|^{2}\Bigr)=\frac{1}{3}\,.\label{eq:MUsum}\end{equation}
Using $\sum_{J}|\bk{\Phi}{J_{z}}|^{2}=1$, i.e. the completeness relation
of the basis $\{\ket{J_{z}}\}$, and $|\bk{\Phi}{0_{z}}|^{2}=1/3$,
we find that $|\bk{\Phi}{1_{z}}|^{2}+|\bk{\Phi}{2_{z}}|^{2}=2/3$.
Substituting this identity into (\ref{eq:MUsum}) leaves us with
$|\bk{\phi}{\hat{u}0_{z}}|^{2}=1/2$,
so that $|\bk{\Phi}{1_{z}}|^{2}=|\bk{\Phi}{2_{z}}|^{2}=1/3$ as well.
A similar argument applied to the pair
$\{\ket{\hat{u}1_{z},\hat{V1_{z}}},\ket{\hat{u}1_{z},\hat{V}2_{z}}\}$
shows that indeed $|\bk{\phi}{\hat{u}1_{z}}|^{2}=1/2$ and
$|\bk{\Phi}{\hat{V}1_{z}}|^{2}=|\bk{\Phi}{\hat{V}2_{z}}|^{2}=1/3$,
which confirms that the state $\ket{\phi,\Phi}$ is of the desired
form. The \emph{converse} direction of the statement is straightforward.

We conjecture Theorem \ref{thm:product-states-MU-to-bases-d=00003D4,6}
to hold for \emph{all} product dimensions $d\equiv pq$, i.e.
$d=4,6,9,10,\ldots$
However, a proof similar to the one for $d=4,6,$ would rely on the
structure of all product bases in composite dimensions $d>6$ -- which
is not known to us.

\section{MU product bases in dimension four\label{sec:MU-product-bases-dim4}}

\subsection{All pairs of MU product
bases\label{sub:All-pairs-ofMUPB-d=00003D4}}

To construct \emph{pairs} of MU product bases in the space $\mathbb{C}^{4}$,
we check all possibilities to form MU pairs of the product bases displayed in Lemma \ref{thm:Any-orthonormal-productd=00003D4} of Sec.
\ref{sub:all-product-bases-d=00003D4}. We find two families of locally inequivalent MU product bases given in Proposition \ref{thm:five-pairs-d=00003D4} below. The derivation of Proposition \ref{thm:five-pairs-d=00003D4} relies on a technique also used for the six-dimensional case presented in Appendix \ref{sec: AppendixBMUpairs-dim6}.

\begin{prop}
\label{thm:five-pairs-d=00003D4} Any pair of MU product bases in
the space $\mathbb{C}^{2}\otimes\mathbb{C}^{2}$ is equivalent to
a member of the families \begin{align}
\mathcal{P}_{0}^{(4)} & \equiv\{\ket{j_{z},k_{z}};\,\ket{j_{x},k_{x}}\}\,, \nonumber \\
\mathcal{P}_{1}^{(4)} &
\equiv\{\ket{0_{z},k_{z}},\ket{1_{z},\hat{s}_{\mu}k_{z}};\,\ket{j_{x},0_{x}},\ket{\hat{r}_{\nu}j_{x},1_{x}}\}\,,\end{align}
where $j,k=0,1$, and the unitary operator
$\hat{r}_{\nu}$ rotates the basis $\{\ket{j_{x}}\}\equiv\{\ket{k_{x}}\}\equiv\{\ket{\pm}\}$
into the $xy$-plane according to $\hat{r}_{\nu}\ket{\pm}=(\ket{0_{z}}\pm
e^{i\nu}\ket{1_{z}})/\sqrt{2}$
for $\nu\in(0,\pi)$; the operator \textup{$\hat{s}_{\mu}$
}\textup{\emph{generates rotations about the $x$-axis, i.e. $\hat{s}_{\mu}\ket{k_z}=(\ket{0_{x}} +(-1)^{k_z} e^{i\mu}\ket{1_{x}})/\sqrt{2}$ for $\mu\in[0,\pi)$}}.
\end{prop}
The pair $\mathcal{P}_{0}^{(4)}$ is the Heisenberg-Weyl pair consisting
of two direct product bases. The pair of MU bases $\mathcal{P}_{1}^{(4)}$
is a \emph{two-parameter} family and may contain direct and indirect product bases. Notice that the operator $\hat{s}_\mu$ can act as the identity since the first basis of $\mathcal{P}_{1}^{(4)}$ may be the standard basis $\{\ket{j_z,k_z}\}$.

The pair $\mathcal{P}_{1}^{(4)}$ turns out to be \emph{equivalent} under \emph{non-local} transformations to the Fourier basis as follows from mapping the first basis to the standard basis $\{\ket{j_{z},k_{z}}\}$. Thus, we have obtained all known pairs of MU bases in dimension four (cf. Sec. 3 of \cite{brierley+10}) in spite of limiting ourselves initially to MU product bases only.

\subsection{All triples of MU product bases}

Now we are in a position to derive all triples of MU product bases
in dimension $d=4$: we need to determine which of the pairs of MU
product bases given in Proposition \ref{thm:five-pairs-d=00003D4} can
be extended by a third MU product basis.

It is easy to see that the MU pair
$\mathcal{P}_{0}^{(4)}\equiv\{\ket{j_{z},k_{z}};\,\ket{j_{x},k_{x}}\}$
can be extended by adjoining a third direct product basis, namely
$\ket{j_{y},k_{y}}$, resulting in the standard Heisenberg-Weyl triple.
This is the \emph{only} possibility, as follows immediately from Theorem
\ref{thm:product-states-MU-to-bases-d=00003D4,6}: a product state
$\ket{\phi,\Phi}$ is MU to both $\{\ket{j_{z},k_{z}}\}$ and
\{$\ket{j_{x},k_{x}}\}$
only if $\ket{\phi}$ is MU both to $\{\ket{j_{z}}\}$ and \{$\ket{j_{x}}\}$,
and if $\ket{\Phi}$ is MU both to $\{\ket{k_{z}}\}$ and \{$\ket{k_{x}}\}$.

The pair $\mathcal{P}_{1}^{(4)}$ of MU bases cannot be extended,
not even by a single MU product state. To extend the
pair by an MU product state, one would need
to find a state in $\mathbb{C}^{2}$ which is MU to the three bases
$\{\ket{k_{z}}\}$, $\{\ket{k_{x}}\}$ and $\{\ket{\hat{r}_{\nu}k_{x}}\}$.
Since $\nu\in(0,\pi)$, no two of these three bases coincide and there is no state in the space $\mathbb{C}^{2}$ simultaneously
MU to three distinct bases. As a consequence, the number of MU product triples is rather limited in dimension four.
\begin{prop}
\label{thm:All-triples-d=00003D4} Any triple of MU product bases
in the space $\mathbb{C}^{2}\otimes\mathbb{C}^{2}$ is equivalent
to \begin{equation}
\mathcal{T}_{0}^{(4)}\equiv\left\{
\ket{j_{z},k_{z}};\,\ket{j_{x},k_{x}};\,\ket{j_{y},k_{y}}\right\}
\,.\end{equation}
\end{prop}
Using Theorem \ref{thm:product-states-MU-to-bases-d=00003D4,6} again,
the non-existence of even a single product state MU to the triple
$\mathcal{T}_{0}^{(4)}$ follows immediately---all states MU to the
triple must be entangled. 

This observation agrees with results reported earlier. For the two-qubit system considered here, a construction of the five MU bases based on the Galois field $GF(4)$ has been given in \cite{klimov+07}. The complete sets obtained turn out to be equivalent under local unitary transformations, and they necessarily consist of three MU bases made up from separable (i.e. product) bases while the remaining two contain maximally entangled states only. This structure also emerges from an approach which exploits the fact that any \emph{complete} set of MU bases of a bipartite system in $\mathbb{C}^d$ contains a \emph{fixed} $d$-dependent amount of entanglement \cite{Wiesniak+11}. When $d=4$, this result implies that for a complete set of MU bases containing the triple $\mathcal{T}_{0}^{(4)}$, the other two bases of the quintuple must consist of \emph{entangled} states -- in fact, only maximally entangled states are permitted. In \cite{Lawrence11}, the entanglement structure of complete sets of MU bases related to Heisenberg-Weyl operators in prime-power dimensions has been studied leading to a generalization of the result for dimension $d=4$: in bipartite systems of dimension $d=p^2$ a number of  $(p+1)$ MU bases must consist of product states while the remaining ones contain only maximally entangled states.

\section{MU product bases in dimension six}

\subsection{All pairs of MU product bases}
\label{MUPBs in d=6}
We will now construct all pairs of MU product bases in dimension six
following the method used in dimension four (cf. Sec.
\ref{sub:all-product-bases-d=00003D4}).
To obtain a MU pair we take each basis listed in Lemma \ref{thm:product-bases-d=00003D6} and go through all possibilities of adding one of the product bases  $\mathcal{B}_{0}$
to $\mathcal{B}_{3}$ (cf. Eqs. (\ref{eq:basis0dim6},\ref{eq:basis1dim6},\ref{eq:basis2dim6},\ref{eq:basis3dim6})
of Appendix \ref{sec:Appendixproductbasesdim6}).

When constructing pairs of MU product bases, it is not necessary
to include the basis $\mathcal{I}_{2}$ in Lemma
\ref{thm:product-bases-d=00003D6}.
We will show now that the operator $\hat{V}$ must either act as the
identity on the pair of states $\{\ket{1_{z}},\ket{2_{z}}\}$ or swap them,
i.e. only $\alpha=0$ or $\beta=0$ are allowed in the expression
$\hat{V}\ket{1_{z}}=\alpha\ket{1_{z}}+\beta\ket{2_{z}}$. However,
in both cases the simplified product basis $\mathbf{\mathcal{I}}_{2}$
turns into a special case of $\mathbf{\mathcal{I}}_{3},$ given in
(\ref{eq:I3}).

Here is the reason why the operator $\hat{V}$ must simplify in the
way just described. Apply Theorem
\ref{thm:product-states-MU-to-bases-d=00003D4,6}
to the product state $\ket{\phi,\Phi}$ required to be MU to
$\mathbf{\mathcal{I}}_{2}$:
the state $\ket{\Phi}$ must be MU to all six vectors
of $\mathbb{C}^{3}$ present in $\mathbf{\mathcal{I}}_{2}$. Consequently,
all states in $\mathbb{C}^{3}$ which occur in the bases $\mathcal{B}_{0}$
to $\mathcal{B}_{3}$, defined in Eqs.
(\ref{eq:basis0dim6},\ref{eq:basis1dim6},\ref{eq:basis2dim6},\ref{eq:basis3dim6}) -- these are all candidates for a second product basis MU to $\mathcal{I}_{2}$ -- must be MU to the standard basis $\{\ket{J_{z}}\}$ of $\mathbb{C}^{3}$.
Now, each of these four bases contains another orthonormal basis of
$\mathbb{C}^{3}$, namely $\{\ket
A,\ket{A^{\perp}},\ket{A^{\perp\!\!\!\perp}}\}$.
There is a two-parameter
family of such states, given in Eq. (\ref{eq:RJx}). However, these
states must also be MU to the state $\hat{V}\ket{1_{z}}$ of the basis
$\mathbf{\mathcal{I}}_{2}$. For the states $\ket A$ and $\ket{A^{\perp}}$,
this requirement reads \begin{equation}
| \bra{A}(\alpha\ket{1_{z}}+\beta\ket{2_{z}})|^{2}=|\bra{A^{\perp}}(\alpha\ket{1_{z}}+\beta\ket{2_{z}})|^{2}=\frac{1}{3}\,.\end{equation}
Now using the explicit expressions of the states $\ket A$ and $\ket{A^{\perp}}$
given in Eq. (\ref{zandxmatrix}) and the identity
$|\alpha|^{2}+|\beta|^{2}=1$,
the first equality leads to \begin{equation}
| 1+\omega|\,|\alpha|\,|\beta|=|\alpha|\,|\beta|\:,\end{equation}
which implies that either $\alpha\equiv0$ or $\beta\equiv0$. Thus,
for the construction of pairs it is sufficient to use the restricted
basis\begin{equation}
\mathcal{I}_{2}^{\prime}=\{\ket{j_{z},0_{z}},\ket{\hat{u}j_{z},1_{z}},\ket{\hat{u}j_{z},2_{z}}\}\,\label{eq:simplifiedI2}\end{equation}
instead of $\mathbf{\mathcal{I}}_{2}$ given in Lemma
\ref{thm:product-bases-d=00003D6}.
All bases of this form, however,
are contained in $\mathcal{I}_{3}$ if one chooses
$\hat{v}=\hat{w}\equiv\hat{u}$
in (\ref{eq:I3}). This simplification also holds for the basis
$\mathcal{B}_{2}$ when
occurring in a pair of product bases.

The actual derivation of all MU product bases in dimension six is lengthy but straightforward. The calculations have been relegated to Appendix \ref{sec: AppendixBMUpairs-dim6} except for the pairing of the basis $\mathcal{I}_{1}$ with $\mathcal{B}_{1}$, which gives rise to the pair $\mathcal{P}_{3}$. The proof that no other (non-trivial) pair of MU product bases results from $\{ \mathcal{I}_{1}; \mathcal{B}_{1} \} $ has been obtained by A. Sudbery, and it is given in Appendix \ref{sec:AppendixUniqueness}. We now summarise the results derived in these two appendices.
\begin{thm}
\label{thm:All-pairs-d=00003D6} Any pair of MU product bases in the
space $\mathbb{C}^{2}\otimes\mathbb{C}^{3}$ is equivalent to a member
of the families \begin{align}
\mathcal{P}_{0} & =\{\ket{j_{z},J_{z}};\,\ket{j_{x},J_{x}}\}\,,
\nonumber \\
\mathcal{P}_{1} &
=\{\ket{j_{z},J_{z}};\,\ket{0_{x},J_{x}},\ket{1_{x},\hat{R}_{\xi,\eta}J_{x}}\}\,, \nonumber \\
\mathcal{P}_{2} &
=\{\ket{0_{z},J_{z}},\ket{1_{z},J_{y}};\,\ket{0_{x},J_{x}},\ket{1_{x},J_{w}}\}\,, \nonumber \\
\mathcal{P}_{3} &
=\{\ket{0_{z},J_{z}},\ket{1_{z},\hat{S}_{\zeta,\chi}J_{z}};\,\ket{j_{x},0_{x}},\ket{\hat{r}_{\sigma}j_{x},1_{x}},\ket{\hat{r}_{\tau}j_{x},2_{x}}\}\,,\end{align}
 with $j=0,1$ and $J=0,1,2$. The unitary operator $\hat{R}_{\xi,\eta}$
is defined as
$\hat{R}_{\xi,\eta}=\kb{0_{z}}{0_{z}}+e^{i\xi}\kb{1_{z}}{1_{z}}+e^{i\eta}\kb{2_{z}}{2_{z}}\,,$
for $\eta,\xi\in[0,2\pi)$, and $\hat{S}_{\zeta,\chi}$ is defined
analogously with respect to the $x$-basis; the unitary operators
$\hat{r}_{\sigma}$ and $\hat{r}_{\tau}$ act on the basis
$\{\ket{j_{x}}\}\equiv\{\ket{\pm}\}$
according to $\hat{r}_{\sigma}\ket{j_{x}}=(\ket{0_{z}}\pm
e^{i\sigma}\ket{1_{z}})/\sqrt{2}$
for $\sigma\in(0,\pi)$, etc.
\end{thm}
As before, the ranges of the parameters are assumed to be such that
no MU product pair occurs more than once in the list. The pairs $\mathcal{P}_{0}$ and $\mathcal{P}_{2}$ have no parameter dependence, the pair $\mathcal{P}_{1}$ depends on two parameters, while $\mathcal{P}_{3}$ is a four-parameter family.

Theorem \ref{thm:All-pairs-d=00003D6} represents the first main result of this paper. It states that there are continuously many possibilities to select pairs of MU bases which, however, can be listed exhaustively. In the remainder of this paper we will proceed by \emph{analytically} constructing all \emph{triples} of MU bases which exist in $d=6$. This will lead to our most important result, namely Theorem 4 in Sec. \ref{sec:excluding-product-bases} which states the impossibility to extend any MU product triple by even a single MU vector. Thus, complete sets of MU bases in $d=6$ will contain at most \emph{pairs} of MU product bases. 

An alternative method to exploit Theorem \ref{thm:All-pairs-d=00003D6} has been pursued in \cite{mcnulty+11}. Upon using suitable \emph{non-local} unitary transformations and known results obtained by computer-algebraic methods, the strongest possible statement about MU product bases is then derived: if a complete set of seven MU bases exists, it will contain at most \emph{one} product basis -- which may be chosen to be the standard basis.

\subsection{All triples of MU product bases}

It is straightforward to enlarge the existing pairs of MU product
bases in Theorem \ref{thm:All-pairs-d=00003D6} to triples: simply
add the MU product bases listed in Lemma \ref{thm:product-bases-d=00003D6},
one after the other, to each of the pairs $\mathcal{P}_{0}$ to
$\mathcal{P}_{3}$ and check whether a valid
MU product triple results.

Neither of the pairs $\mathcal{P}_{2}$ and $\mathcal{P}_{3}$ in Theorem \ref{thm:All-pairs-d=00003D6}
can be extended by a single MU product state. To do so, we would need
a vector MU to the three distinct bases $\{\ket{j_z}\}$, $\{\ket{j_x}\}$ and $\{\ket{\hat{r}_\sigma j_x}\}$ in the space $\mathbb{C}^{2}$, or a vector mutually unbiased to four MU bases in the space $\mathbb{C}^{3}$. No such states exist, implying that any state mutually unbiased to these pairs must be entangled.

The pairs $\mathcal{P}_{0}$ and $\mathcal{P}_{1}$ \emph{can}
be extended by a further MU product basis since there exist vectors of the spaces $\mathbb{C}^2$ and $\mathbb{C}^3$ that satisfy the necessary conditions. To obtain the complete list of all MU product triples in $\mathbb{C}^{6}$ we thus need to search for possible extensions of these two pairs by a third product basis. Starting with $\mathcal{P}_{0}$,
it is possible to extend this pair by either $\mathcal{B}_{0}$ or
$\mathcal{B}_{1}$.

$\bullet\,\{\mathcal{P}_{0};\mathcal{B}_{0}\}$: If we choose
the third basis to be of the form $\mathcal{B}_{0}$, there are only
two choices, $\{\ket{j_{y},J_{y}}\}$ or $\{\ket{j_{y},J_{w}}\}$.
Using the local complex conjugation $\hat{I}\otimes\hat{K}$, the
resulting triples are found to be equivalent, \begin{equation}
\{\ket{j_{z},J_{z}};\,\ket{j_{x},J_{x}};\,\ket{j_{y},J_{y}}\}\sim\{\ket{j_{z},J_{z}};\,\ket{j_{x},J_{x}};\,\ket{j_{y},J_{w}}\}\,;\end{equation}
consequently, all triples of this type are equivalent to the Heisenberg-Weyl
triple \begin{equation}
\mathcal{T}_{0}=\{\ket{j_{z},J_{z}};\,\ket{j_{x},J_{x}};\,\ket{j_{y},J_{y}}\}\,.\end{equation}
 $\bullet\,\{\mathcal{P}_{0};\mathcal{B}_{1}\}$: If we extend
$\mathcal{P}_{0}$ by an indirect product basis of the form
$\mathcal{B}_{1}$, there are only two choices,
$\{\ket{0_{y},J_{y}},\ket{1_{y},J_{w}}\}$
or $\{\ket{0_{y},J_{w}},\ket{1_{y},J_{y}}\}$. Again, a local complex
conjugation $\hat{k}\otimes\hat{I}$ maps one of the triples into the
other, \begin{equation}
\{\ket{j_{z},J_{z}};\ket{j_{x},J_{x}};\,\ket{0_{y},J_{y}},\ket{1_{y},J_{w}}\}\sim\{\ket{j_{z},J_{z}};\,\ket{j_{x},J_{x}};\,\ket{0_{y},J_{w}},\ket{1_{y},J_{y}}\}\,,\end{equation}
leaving us with the triple
\begin{equation}
\mathcal{T}_{1}=\{\ket{j_{z},J_{z}};\,\ket{j_{x},J_{x}};\,\ket{0_{y},J_{y}},\ket{1_{y},J_{w}}\}\,.\end{equation}
Now turning to the pair $\mathcal{P}_{1}$, we again attempt
to obtain a triple by adding either $\mathcal{B}_{0}$ or $\mathcal{B}_{1}$.

$\bullet\,\{\mathcal{P}_{1};\mathcal{B}_{0}\}$ or
$\{\mathcal{P}_{1};\mathcal{B}_{1}\}$:
First, extend the pair $\mathcal{P}_{1}$ by a direct product
basis, resulting in either $\{\ket{j_{z},J_{z}};\ket{0_{x},J_{x}},\ket{1_{x},J_{y}};\ket{j_{y},J_{w}}\}$
or $\{\ket{j_{z},J_{z}};\ket{0_{x},J_{x}},\ket{1_{x},J_{w}};\ket{j_{y},J_{y}}\}$.
It is not difficult to apply suitable LETs to transform them into
the triple $\mathcal{T}_{1}$. Now extend the pair $\mathcal{P}_{1}$ by an indirect product basis $\mathcal{B}_1$. This leads to a contradiction since we would need the states $\{\ket{\hat{R}_{\xi,\eta}J_{x}}\}$ in $\mathcal{P}_1$ to coincide with $\{\ket{J_{x}}\}$, which is not allowed.

This completes the construction of all MU product triples in dimension
six, leading to the second main result of this paper.
\begin{thm}
\label{thm:two-triples-d=00003D6} Any triple of MU product bases
in the space $\mathbb{C}^{2}\otimes\mathbb{C}^{3}$ is equivalent
to either  
\begin{align}
\mathcal{T}_{0} &
=\{\ket{j_{z},J_{z}};\,\ket{j_{x},J_{x}};\,\ket{j_{y},J_{y}}\}\,,
 \nonumber \\
\mbox{or } \mathcal{T}_{1} &
=\{\ket{j_{z},J_{z}};\,\ket{j_{x},J_{x}};\,\ket{0_{y},J_{y}},\ket{1_{y},J_{w}}\}\,.\end{align}
\end{thm}
According to Theorem \ref{thm:product-states-MU-to-bases-d=00003D4,6},
neither of these triples can be extended by a single MU product state.
Thus, any complete set of seven MU bases in dimension six will contain
at most three product bases, and if it does, the triple must be equivalent
to one of those in Theorem \ref{thm:two-triples-d=00003D6}. In the following section we will obtain an even stronger result.

\section{Excluding triples of MU product bases from complete sets}
\label{sec:excluding-product-bases}

In this section we derive the third main result of this paper.
\begin{thm}\label{thm:noproductbases}
No triple of MU product bases in dimension six can be extended by a single  MU vector. 
\end{thm}
In other words, no complete set of seven MU bases in $d=6$ contains a triple of MU product bases. This result relies on a computer-algebraic proof in \cite{grassl+04}, which finds a total of 48 vectors MU to the pair of eigenbases of the Heisenberg-Weyl operators $X_6$ and $Z_6$, giving rise to sixteen different orthonormal bases. However, none of these bases allows one to extend the given pair beyond a triple of MU bases.

The present construction of MU product triples effectively produces
twelve (and only twelve) product vectors that are MU to the pair $\mathcal{P}_{0}=\{\ket{j_{z},J_{z}};\,\ket{j_{x},J_{x}}\}$, namely $\{\ket{j_{y},J_{y}}\}$ and
$\{\ket{j_{y},J_{w}}$\}, and they give rise to the only two inequivalent
triples of MU bases, $\mathcal{T}_{0}$ and $\mathcal{T}_{1}$.
Since $\mathcal{P}_0$ is equivalent to the eigenbases of $X_6$ and $Z_6$, clearly these twelve product vectors must figure among the 48 vectors given
in \cite{grassl+04}.

To show this, we must first deal with a difference in our definition of the HW operators. The HW pair used in \cite{grassl+04} does not have the same form as $\mathcal{P}_0$ since the $x$-basis in \cite{grassl+04} is the eigenbasis of the operator $X_6$, whereas we have used the eigenbasis of the operator $X_2\otimes X_3$ (cf. Eq. (\ref{HWoperators})). Nevertheless, both pairs of bases turn out to be equivalent using a \emph{non-local} unitary transformation. By writing the operators as matrices, we find that $X_2\otimes X_3=P_{25}X_6P_{25}$, where $P_{25}$ is a permutation matrix permuting rows two and five. This non-local transformation brings the eigenbasis of $X_6$ into product form, i.e. $\{\ket{j_x,J_x}\}$, by multiplying it with $P_{25}$ from the left. 

The same transformation must also be applied to the list of $48$ vectors so that they are MU to the pair $\mathcal{P}_0$. After multiplying each of these vectors by the matrix $P_{25}$ from the left, one easily identifies the twelve product vectors, numbered by $1,2,5,6,9,10,13,14,17,18,21$ and $22$ in the Appendix of the updated version of \cite{grassl+04}. For example, the vector labelled (1) transforms as follows:
\begin{align}
\frac{1}{\sqrt{6}}P_{25}(1,\alpha^5,1,-\alpha^3,-\alpha^2,-\alpha^3)^{\text{T}}=&\frac{1}{\sqrt{6}}(1,-\alpha^2,1,-\alpha^3,\alpha^5,-\alpha^3)^{\text{T}} \nonumber \\
\equiv&\frac{1}{\sqrt{6}}(1,\omega^2,1,-i,-i\omega^2,-i)^{\text{T}}\nonumber \\
\equiv&\frac{1}{\sqrt{6}}(1,-i)^\text{T}\otimes(1,\omega^2,1)^{\text{T}}
 \end{align}
where $\alpha=e^{2\pi i/12}$ and $\omega=e^{2\pi i/3}$. This vector is the product state $\ket{1_y,1_y}$.

The twelve vectors give rise to four of the sixteen orthonormal bases which are MU to the original pair. These product bases are covered by the product bases we construct when extending the Heisenberg-Weyl pair $\mathcal{P}_{0}$ to a triple; however, only two of the four triples are locally inequivalent as follows from exploiting suitable local equivalence transformations.

Upon combining the computer-aided result just described with Theorem
\ref{thm:two-triples-d=00003D6} it is straightforward to arrive at a result that excludes product triples from being part of a complete set of seven MU bases. The triples of MU product bases $\mathcal{T}_0$ and $\mathcal{T}_1$ both contain the Heisenberg-Weyl pair $\mathcal{P}_{0}=\{\ket{j_{z},J_{z}};\,\ket{j_{x},J_{x}}\}$,
and it is impossible to extend this pair by more than a single MU basis according to \cite{grassl+04}. Since Theorem \ref{thm:two-triples-d=00003D6} provides an \emph{exhaustive} list of MU triples in the space $\mathbb{C}^2\otimes\mathbb{C}^3$, it follows that \emph{no complete set of seven MU bases in $d=6$ contains a triple of MU product bases}.

\section{Summary and discussion}

By limiting ourselves to orthonormal product bases, we have been able
to obtain a number of analytic results regarding the existence of MU
bases of the space $\mathbb{C}^{2}\otimes\mathbb{C}^{3}$. After identifying
all orthonormal product bases of this space, presented in Lemma
\ref{thm:product-bases-d=00003D6},
we have constructed an exhaustive list of pairs of MU product bases. They come in four different flavours according to Theorem
\ref{thm:All-pairs-d=00003D6}. Next, Theorem \ref{thm:two-triples-d=00003D6} states that, in addition to the Heisenberg-Weyl triple, there is only one other locally inequivalent triple of MU product bases. The absence of quadruples of MU product bases agrees with Zauner's conjecture \cite{zauner+99} that there are no more than three MU bases in dimension six.

The derivation of the list of MU product pairs and triples has been
simplified considerably by the content of Theorem \ref{thm:product-states-MU-to-bases-d=00003D4,6}. It spells out severe restrictions on the form of product states required to be MU to certain sets of orthonormal states in the space
$\mathbb{C}^{2}\otimes\mathbb{C}^{3}$.
We have established Theorem \ref{thm:product-states-MU-to-bases-d=00003D4,6}
for dimensions $d=4$ and $d=6$ only, since the proof relies on enumerating
all orthonormal product bases in these dimensions.

Theorem \ref{thm:two-triples-d=00003D6} allows us to partly
replicate results obtained by means of a computer-algebraic method. Out of the 48 vectors mutually unbiased to the Heisenberg-Weyl pair $\mathcal{P}_0$, found in \cite{grassl+04}, we successfully recover twelve, and they are shown to be equivalent to product vectors.

The most important consequence of exhaustively enumerating MU product bases in dimension six is a bound on their allowed number in complete sets of MU bases. Applying Theorem \ref{thm:product-states-MU-to-bases-d=00003D4,6} to the triples of MU product bases in  $\mathbb{C}^{2}\otimes\mathbb{C}^{3}$, namely $\mathcal{T}_{0}$ and $\mathcal{T}_{1}$, directly implies that no single \emph{product} state can be MU to any of them. However, a stronger result is within reach, spelled out in Theorem \ref{thm:noproductbases}: it is impossible to complement either $\mathcal{T}_{0}$ or $\mathcal{T}_{1}$ by \emph{any} MU vector. This follows from combining 
Theorem \ref{thm:two-triples-d=00003D6} with the results derived in \cite{grassl+04}. Thus, \emph{a complete set of MU bases in dimension six cannot contain a product triple}. This is in marked contrast to the prime-power dimension $p^2$ where a complete set of MU bases necessarily contains $(p+1)$ MU product bases constructed from  the tensor products of Heisenberg-Weyl operators \cite{Lawrence11}. The exhaustive list of MU product pairs given in Theorem 2 has been used to derive a result even stronger than Theorem 4, reducing the number of allowed MU product bases to just a single one \cite{mcnulty+11}.

A similar situation has been described in \cite{aschbacher+07} where a different class of MU bases is studied. Given a ``nice unitary error basis'', consisting of $d^2$ suitable matrices, one can search for MU bases within these sets. In the case of dimension six, it is shown that any partition of a nice error basis gives rise to no more than three MU bases. This limitation and the non-existence of more that three MU \emph{product} bases are independent results: MU product bases and MU bases arising from nice unitary error bases are structurally different. For example, our construction reproduces the continuous family $\mathcal{P}_1^{(4)}$ of MU product pairs in $d=4$, and it is known that some of the pairs in this family are \emph{inequivalent} to MU bases stemming from nice unitary error bases \cite{Klappenecker+05}.

Our considerations are backed by deriving corresponding results in the Hilbert space of two qubits, i.e. $\mathbb{C}^{2}\otimes\mathbb{C}^{2}$.
In this case there exists a symmetry between the two factors and the enumeration of MU product pairs and triples is much simpler. Clearly, when a qubit is combined with a qutrit, no such symmetry exists. We believe that the symmetries between the subsystems present in only prime-power dimensions
are the ultimate reason that additional ``identities'' exist which
allow for the construction of \emph{complete} sets of MU bases.

Let us conclude by formulating a conjecture which emerges naturally from our
results: we expect Theorem \ref{thm:product-states-MU-to-bases-d=00003D4,6}
to hold for \emph{all} composite dimensions $d=pq\geq4$, not only for $d=4$ and $d=6$. Our pedestrian proof in these dimensions relies on enumerating
all orthonormal product bases. However, the set of product bases in
composite dimensions is likely to possess a certain structure which,
once spelled out, should allow for a more elegant proof applicable
to arbitrary composite dimensions.

\subsubsection*{Acknowledgements}
The proof of Theorem \ref{tony}, presented in Appendix \ref{sec:AppendixUniqueness}, has been found by A. Sudbery; we gratefully acknowledge his permission to reproduce it here. We thank S. Brierley, M. Grassl, and A. Sudbery for comments and suggestions. This work has been supported by EPSRC.

\appendix

\section{Appendix \label{sec:Appendixproductbasesdim6}}

In this Appendix we derive all product bases in dimension six reported
in Lemma \ref{thm:product-bases-d=00003D6}. The six states
$\ket{\psi_{j}}\,,j=1\ldots6$, of a product basis $\{\ket{\psi_{j},\Psi_{j}}\}$,
defined in (\ref{eq: candidatebasisindim6}) must contain at
least \emph{two} (possibly identical) pairs of orthogonal states. If there was only one
pair (with the remaining four states of the space $\mathbb{C}^2$ non-orthogonal), the orthogonality conditions (\ref{eq:orthoconddim6}) would require \emph{four} orthogonal states $\ket{\Psi_{j}}\in\mathbb{C}^{3}$, which do not exist. Thus, denoting the orthogonal pairs by $\{\ket a,\ket{a^{\perp}}\}$ and $\{\ket b,\ket{b^{\perp}}\}$, the product basis must take the form 
\begin{equation}
\Bigl\{\ket{a,\Psi_{1}},\,\ket{a^{\perp},\Psi_{2}},\,\ket{b,\Psi_{3}},\,\ket{b^{\perp},\Psi_{4}},\,\ket{\psi_{5},\Psi_{5}},\,\ket{\psi_{6},\Psi_{6}}\Bigr\}\,.\end{equation}
The states $\ket{\psi_{5}}$ and $\ket{\psi_{6}}$ must also be an
orthogonal pair. To see this, assume that they are skew (or identical) \emph{and}
they are both skew to the states $\ket a$ and $\ket b$; then the
state $\ket{\Psi_{6}}$, for example, must be orthogonal to the orthonormal
triple $\{\ket{\Psi_{1}},\ket{\Psi_{3}},\ket{\Psi_{5}}\}$, which is
impossible. Here we have assumed that $\ket{a}$ and $\ket{b}$ are \emph{not} orthogonal; if they are, we use the orthonormal triple $\{\ket{\Psi_{1}},\ket{\Psi_{4}},\ket{\Psi_{5}}\}$ instead. The same conclusion can be drawn if the states $\ket{\psi_{5}}$
and $\ket{\psi_{6}}$ are skew (or identical) but one of them coincides with any
of the four states $\ket a$, $\ket b$, $\ket{a^{\perp}}$ or $\ket{b^{\perp}}$.
Thus, we conclude that any product basis of the space $\mathbb{C}^{6}$
must be of the form\begin{equation}
\Bigl\{\ket{a,\Psi_{1}},\,\ket{a^{\perp},\Psi_{2}},\,\ket{b,\Psi_{3}},\,\ket{b^{\perp},\Psi_{4}},\,\ket{c,\Psi_{5}},\,\ket{c^{\perp},\Psi_{6}}\Bigr\}\,.\label{eq:
threeBasesofC2}\end{equation}
Now it is obvious that we need to consider three different possibilities
depending on how many of the three bases of the space $\mathbb{C}^{2}$
coincide.

\textbf{Case 1}: If all three bases coincide, we have\begin{equation}
\Bigl\{\ket{a,\Psi_{1}},\,\ket{a^{\perp},\Psi_{2}},\,\ket{a,\Psi_{3}},\,\ket{a^{\perp},\Psi_{4}},\,\ket{a,\Psi_{5}},\,\ket{a^{\perp},\Psi_{6}}\Bigr\}\,.\end{equation}
These six states are orthogonal only if the three
states $\ket{\Psi_{1}},\ket{\Psi_{3}}$ and $\ket{\Psi_{5}}$ are
orthogonal to each other, as well as the triple
$\{\ket{\Psi_{2}},\ket{\Psi_{4}},\ket{\Psi_{6}}\}$.
Upon denoting the first triple by $\{\ket
A,\ket{A^{\perp}},\ket{A^{\perp\!\!\!\perp}}\}$,
where $\ket{A^{\perp\!\!\!\perp}}\in\mathbb{C}^{3}$ is a vector orthogonal
to $\ket A$ and $\ket{A^{\perp}}$, we obtain\begin{equation}
\mathcal{B}_{1}=\Bigl\{\ket{a,A},\,\ket{a,A^{\perp}},\,\ket{a,A^{\perp\!\!\!\perp}},\,\ket{a^{\perp},B},\,\ket{a^{\perp},B^{\perp}},\,\ket{a^{\perp},B^{\perp\!\!\!\perp}}\Bigr\}\,,\label{eq:basis1dim6}\end{equation}
also introducing an arbitrary second triple of orthogonal states.
If the two triples coincide, we find the important special case of
a direct product basis\begin{equation}
\mathcal{B}_{0}=\Bigl\{\ket{a,A},\,\ket{a,A^{\perp}},\,\ket{a,A^{\perp\!\!\!\perp}},\,\ket{a^{\perp},A},\,\ket{a^{\perp},A^{\perp}},\,\ket{a^{\perp},A^{\perp\!\!\!\perp}}\Bigr\}\,,\label{eq:basis0dim6}\end{equation}

\textbf{Case 2}: If only two of the bases in $\mathbb{C}^{2}$ coincide,
we find\begin{equation}
\Bigl\{\ket{a,\Psi_{1}},\,\ket{a^{\perp},\Psi_{2}},\,\ket{b,\Psi_{3}},\,\ket{b^{\perp},\Psi_{4}},\,\ket{b,\Psi_{5}},\,\ket{b^{\perp},\Psi_{6}}\Bigr\}\,.\end{equation}
As in Case 1, each of the triples
$\{\ket{\Psi_{1}},\ket{\Psi_{3}},\ket{\Psi_{5}}\}$
and $\{\ket{\Psi_{2}},\ket{\Psi_{4}},\ket{\Psi_{6}}\}$ must be an
orthonormal basis of $\mathbb{C}^{3}$. However, we also need to
have\begin{equation}
\bk{\Psi_{1}}{\Psi_{4}}=\bk{\Psi_{1}}{\Psi_{6}}=0\,,\end{equation}
which means that $\ket{\Psi_{3}}$ and $\ket{\Psi_{5}}$ span the
same subspace as $\ket{\Psi_{4}}$ and $\ket{\Psi_{6}}$. It follows
that $\ket{\Psi_{1}}\equiv\ket{\Psi_{2}}$; upon calling this state
$\ket A$, we are led to a new class of product bases of $\mathbb{C}^{6}$
given by\begin{equation}
\mathcal{B}_{2}=\Bigl\{\ket{a,A},\,\ket{a^{\perp},A},\,\ket{b,A^{\perp}},\,\ket{b^{\perp},\hat{V}A^{\perp}},\,\ket{b,A^{\perp\!\!\!\perp}},\,\ket{b^{\perp},\hat{V}A^{\perp\!\!\!\perp}}\Bigr\}\,,\label{eq:basis2dim6}\end{equation}
where
$\hat{V}\ket{A^{\perp}}=\alpha\ket{A^{\perp}}+\beta\ket{A^{\perp\!\!\!\perp}}$
and
$\hat{V}\ket{A^{\perp\!\!\!\perp}}=\bar{\beta}\ket{A^{\perp}}-\bar{\alpha}\ket{A^{\perp\!\!\!\perp}}$,
i.e. $\hat{V}$ is any unitary transformation of the two-dimensional
subspace of $\mathbb{C}^{3}$ orthogonal to the state $\ket A$.

\textbf{Case 3}: Finally, we consider the case where the three bases
of the space $\mathbb{C}^{2}$ present in (\ref{eq: threeBasesofC2})
are all different, meaning that $\ket a$, $\ket b$ , and $\ket c$
are pairwise skew. Then, the orthogonality conditions directly imply
that the triples $\{\ket{\Psi_{1}},\ket{\Psi_{3}},\ket{\Psi_{5}}\}$
and $\{\ket{\Psi_{2}},\ket{\Psi_{4}},\ket{\Psi_{6}}\}$ of orthogonal
states must coincide. This leaves us with bases of the form\begin{equation}
\mathcal{B}_{3}=\Bigl\{\ket{a,A},\,\ket{a^{\perp},A},\,\ket{b,A^{\perp}},\,\ket{b^{\perp},A^{\perp}},\,\ket{c,A^{\perp\!\!\!\perp}},\,\ket{c^{\perp},A^{\perp\!\!\!\perp}}\Bigr\}\,.\label{eq:basis3dim6}\end{equation}
These three cases complete the construction of all product bases in
dimension six. Using local equivalence transformations in analogy
to the procedure used in Sec. \ref{sub:all-product-bases-d=00003D4},
one can write the four sets of product bases as displayed in Lemma
\ref{thm:product-bases-d=00003D6}.

\section{Appendix \label{sec: AppendixBMUpairs-dim6}}

In this Appendix we derive all pairs of MU product bases in dimension
six by pairwise combining the orthonormal product bases $\mathcal{B}_{0}$ to $\mathcal{B}_{3}$, defined in Eqs.
(\ref{eq:basis0dim6},\ref{eq:basis1dim6},\ref{eq:basis2dim6},\ref{eq:basis3dim6}).
In principle, we need to look at only 10 of the 16 pairs
$\{\mathcal{B}_{i};\mathcal{B}_{j}\},\, i,j=0\ldots3$,
since the order of the bases does not matter: the pairs
$\{\mathcal{B}_{i};\mathcal{B}_{j}\}$
and $\{\mathcal{B}_{j};\mathcal{B}_{i}\}$ are equivalent for $i\neq j$.
Using local equivalence transformations, each pair can be brought
to the form $\{\mathcal{I}_{i};\mathcal{B}_{j}\},\, i\leq j$, where
the bases $\mathcal{I}_{0}$ to $\mathcal{I}_{3}$ are those listed
in Lemma \ref{thm:product-bases-d=00003D6}. As shown in the main
text, it is not actually necessary to consider the bases $\mathcal{I}_{2}$
and $\mathbf{\mathcal{B}}_{2}$ at all, reducing the number of cases
to six. Parameter ranges are assumed so that no pair occurs more than once.

$\bullet\,\{\mathcal{I}_{0};\mathcal{B}_{0}\}$: First we extend $\mathcal{I}_{0}$
to a pair of MU bases by combining it with\begin{equation}
\mathcal{B}_{0}=\Bigl\{\ket{a,A},\,\ket{a,A^{\perp}},\,\ket{a,A^{\perp\!\!\!\perp}},\,\ket{a^{\perp},A},\,\ket{a^{\perp},A^{\perp}},\,\ket{a^{\perp},A^{\perp\!\!\!\perp}}\Bigr\}\,.\end{equation}
The states of $ \mathcal{B}_{0}$ are MU to those of the basis $\mathcal{I}_{0}$ if the
pair of states $\{\ket a,\ket{a^{\perp}}\}$ is any basis of $\mathbb{C}^{2}$
associated with opposite points on the Bloch sphere, i.e. $\ket
a=(\ket{0_{z}}+e^{i\mu}\ket{1_{z}})/\sqrt{2}$
etc., and if the orthonormal basis $\{\ket
A,\ket{A^{\perp}},\ket{A^{\perp\!\!\!\perp}}\}$
is defined as in Eqs. (\ref{eq:RJx}) of Sec. \ref{sub:All-MU-basesdim2and3}.

A local transformation allows us to rotate the states $\{\ket{a},\ket{a^\perp}\}$
into $\{\ket{j_{x}}\}$ and to simultaneously change the basis $\{\ket
A,\ket{A^{\perp}},\ket{A^{\perp\!\!\!\perp}}\}$
into the basis $\{\ket{J_{x}}\}$ of $\mathbb{C}^{3}$ so that we
end up with the known Heisenberg-Weyl MU pair of direct product bases,
\begin{equation}
\mathcal{P}_{0}\equiv\{\ket{j_{z},J_{z}};\,\ket{j_{x},J_{x}}\}\,.\end{equation}

$\bullet\,\{\mathcal{I}_{0};\mathcal{B}_{1}\}$ and $\{\mathcal{I}_{0};\mathcal{B}_{3}\}$: These cases will be covered by the pairs $\{\mathcal{I}_1;\mathcal{B}_1\}$ and $\{\mathcal{I}_1;\mathcal{B}_3\}$, respectively, since we can treat the basis $\mathcal{I}_0$ as a subset of $\mathcal{I}_1$.  

We now construct the three pairs of indirect product bases that contain
$\mathcal{I}_{1}=\{\ket{0_{z},J_{z}},\ket{1_{z},\hat{U}J_{z}}\}$
as the first basis, where the unitary $\hat{U}$ maps the basis
$\{\ket{J_{z}}\}$
of the space $\mathbb{C}^{3}$ to another basis.

$\bullet\,\{\mathcal{I}_{1};\mathcal{B}_{1}\}$: In a first step, we act with
a local unitary on the second basis\begin{equation}
\mathcal{B}_{1}=\Bigl\{\ket{a,A},\,\ket{a,A^{\perp}},\,\ket{a,A^{\perp\!\!\!\perp}},\,\ket{a^{\perp},B},\,\ket{a^{\perp},B^{\perp}},\,\ket{a^{\perp},B^{\perp\!\!\!\perp}}\Bigr\}\end{equation}
to rotate the $a$-basis of states that are MU to $\{\ket{j_{z}}\}$
into the basis $\{\ket{j_{x}}\}$ while the $A$-basis turns into
$\{\ket{J_{x}}\}$, as before. This maps $\mathcal{B}_{1}$ to\begin{equation}
\Bigl\{\ket{0_{x},J_{x}},\,\ket{1_{x},\hat{U}^{\prime}J_{x}}\Bigr\}\,,\end{equation}
where we have introduced a unitary $\hat{U}^{\prime}$ which parameterises
all orthonormal bases of $\mathbb{C}^{3}$ relative to the 
$x$-basis.
The requirement that the states of the pair $\{\mathcal{I}_{1};\mathcal{B}_{1}\}$
be MU now turns into the problem of identifying all pairs of orthonormal
bases of $\mathbb{C}^{3}$, namely  $\{\ket{J_{z}};\,\ket{\hat{U}J_{z}}\}$ and
$\{\ket{J_{x}};\,\ket{\hat{U}^{\prime}J_{x}}\}$,
such that all states of one set are MU to those of the other,
viz.\begin{equation}
| \bk{J_{z}}{\hat{U^{\prime}}J_{x}}|^{2}=|\bk{\hat{U}J_{z}}{J_{x}}|^{2}=|\bk{\hat{U}J_{z}}{\hat{U^{\prime}}J_{x}}|^{2}=\frac{1}{3}\,,\label{eq:I_1B_1condition}\end{equation}
while $|\bk{J_{z}}{J_{x}}|^{2}=1/3$ holds by construction. It is easy to see
that these conditions are satisfied if the bases in (at least) one pair
coincide or all four are different, i.e. they use up a complete set of MU bases
in $\mathbb{C}^{3}$. In Appendix \ref{sec:AppendixUniqueness} we present a proof, due to A. Sudbery, that these are the \emph{only} solutions of the constraints (\ref{eq:I_1B_1condition}). Thus, if $\mathcal{I}_1$ is the standard basis $\{\ket{j_z,J_z}\}$, then we obtain the MU product pair
\begin{equation}
\mathcal{P}_{1}=\{\ket{j_{z},J_{z}};\,\ket{0_{x},J_{x}},\ket{1_{x},\hat{R}_{\xi,\eta}J_{x}}\}\,,
\end{equation}
with $\{\ket{\hat{R}_{\xi,\eta}J_x}\}$ defined in Eq. $(\ref{eq:RJx})$. However, if we use the complete set of MU bases in $\mathbb{C}^3$ we obtain the MU product pair
\begin{equation}
\mathcal{P}_{2}=\{\ket{0_{z},J_{z}},\ket{1_{z},J_{y}};\,\ket{0_{x},J_{x}},\ket{1_{x},J_{w}}\}\,.
\end{equation}

$\bullet\,\{\mathcal{I}_{1};\mathcal{B}_{3}\}$: The second basis reads explicitly
\begin{equation}
\mathcal{B}_{3}=\Bigl\{\ket{a,A},\,\ket{a^{\perp},A},\,\ket{b,A^{\perp}},\,\ket{b^{\perp},A^{\perp}},\,\ket{c,A^{\perp\!\!\!\perp}},\,\ket{c^{\perp},A^{\perp\!\!\!\perp}}\Bigr\}\,,\end{equation}
and suitable LETs map it to\begin{equation}
\left\{
\ket{j_{x},0_{x}},\ket{\hat{r}_{\sigma}j_{x},1_{x}},\ket{\hat{r}_{\tau}j_{x},2_{x}}\right\}
\,,\end{equation}
which involve two rotations of the basis $\{\ket{j_x}\}$ about the $z$-axis, $\hat{r}_\sigma$ and $\hat{r}_\tau$. The operator
$\hat{U}$ in $\mathcal{I}_{1}$ must be chosen such that
$\{\ket{\hat{U}J_{z}}\}$
is MU to the $x$-basis. All such $U(3)$-rotations are given by 
the two-parameter family 
\begin{equation} \label{S-diagonal}
\hat{S}_{\zeta,\chi}=\kb{0_{x}}{0_{x}}+e^{i\zeta}\kb{1_{x}}{1_{x}}+e^{i\chi}\kb{2_{x}}{2_{x}}\,,\end{equation}
 diagonal
in the $x$-basis, and defined in analogy to $\hat{R}_{\xi,\eta}$ in Eq. (\ref{eq:RJx}). Altogether, we obtain a four-parameter family of MU product pairs, 
\begin{equation}
\mathcal{P}_{3}=\{\ket{0_{z},J_{z}},\ket{1_{z},\hat{S}_{\zeta,\chi}J_{z}};\,\ket{j_{x},0_{x}},\ket{\hat{r}_{\sigma}j_{x},1_{x}},\ket{\hat{r}_{\tau}j_{x},2_{x}}\}\,.
\end{equation}

$\bullet\,\{\mathcal{I}_{3};\mathcal{B}_{3}\}$: No pair results when we combine
the product basis $\mathcal{B}_{3}$ with $\mathcal{I}_{3}$. The
standard transformations to simplify $\mathbf{\mathcal{B}}_{3}$ lead
to \begin{equation}
\Bigl\{\ket{j_{x},0_{x}},\,\ket{\hat{r}_{\sigma}j_{x},1_{x}},\,\ket{\hat{r}_{\tau}j_{x},2_{x}}\Bigr\}\,,\end{equation}
since both the $b$-basis and the $c$-basis must be MU to the standard
basis. The only basis MU to the three bases $\{\ket{j_{x}}\}$,
$\{\ket{\hat{r}_{\sigma}j_{x}}\}$,
and $\{\ket{\hat{r}_{\tau}j_{x}}\},$ is the standard basis $\{\ket{j_{z}}\}$, which is also true for the case $\{\ket{\hat{r}_{\sigma}j_{x}}\}=\{\ket{\hat{r}_{\tau}j_{x}}\}$. Consequently, this would force the operators $\hat{u}$ and $\hat{v}$ to be the identity, in contradiction to the assumption that the three bases
of $\mathbb{C}^{2}$ present in $\mathcal{I}_{3}$ do not coincide.

\section{Appendix \label{sec:AppendixUniqueness}}

Here we report a proof by A. Sudbery that the conditions of Eq. (\ref{eq:I_1B_1condition}) in Appendix \ref{sec: AppendixBMUpairs-dim6} are only satisfied if the bases in (at least) one pair coincide or all four bases are mutually unbiased. If $\mathcal{B}_1$ and $\mathcal{B}_2$  are orthonormal bases, we write $\mathcal{B}_1\,\mu\, \mathcal{B}_2$ to mean {\textquotedblleft$\mathcal{B}_1$ and $\mathcal{B}_2$ are mutually unbiased\textquotedblright}.

\begin{thm}
Suppose $\mathcal{B}_0,\mathcal{B}_1,\mathcal{B}_2,\mathcal{B}_3$ are orthonormal bases of $\C^3$ satisfying 
\[ \{\mathcal{B}_0,\mathcal{B}_1\}\hspace{0.2cm}\mu\hspace{0.2cm}\{\mathcal{B}_2,\mathcal{B}_3\}.\]
Then either $\mathcal{B}_0$ and $\mathcal{B}_1$ are equivalent bases or $\mathcal{B}_2$ and $\mathcal{B}_3$ are equivalent bases or all four bases are mutually unbiased.
\label{tony}\end{thm}

Let $\mathcal{B}_0,\mathcal{B}_1,\mathcal{B}_2,\mathcal{B}_3$ be represented by unitary matrices $I,U,V,W$, respectively, where we have chosen $\mathcal{B}_0$ to be the standard basis of $\C^3$. We regard the bases $U$, $UP$ and $UD$, with $P$ a permutation matrix and $D$ a diagonal, as equivalent bases. Note that if two orthonormal bases in $\mathbb{C}^3$, represented by unitary matrices $U$ and $V$, are mutually unbiased, then $U^\dagger V$ (where the dagger denotes hermitian conjugation) is a complex Hadamard matrix $H$. We can write any $(3 \times 3)$ Hadamard matrix as
\be
H = DFD' \text{ or } DF^\dagger D'
\ee
where $D$ and $D'$ are diagonal and $\displaystyle F \equiv F_3$ is the Fourier matrix defined in Eq. (\ref{zandxmatrix}).

The condition $\mathcal{B}_2\,\mu\, \mathcal{B}_0$ implies the unitary $V$ is a Hadamard matrix, and since $F^\dagger = FP$, the basis $\mathcal{B}_2$ is equivalent to a basis represented by $V = DF$. Similarly, $\mathcal{B}_3$ is equivalent to a basis represented by $W = D'F$ where $D'$ is diagonal. Now 
\begin{align}
\mathcal{B}_2\,\mu\, \mathcal{B}_1 \quad &\impl \quad V^\dagger U = KF^{(1)} L,\\[10pt]
\mathcal{B}_3\,\mu\, \mathcal{B}_1 \quad &\impl \quad W^\dagger U = K'F^{(2)} L'
\end{align}
where $K$, $L$, $K'$ and $L'$ are diagonal and $F^{(i)}$ is either $F$ or $F^\dagger$ ($i = 1,2$). Hence
\be\label{W}
U = DFKF^{(1)}L = D'FK'F^{(2)}L'.
\ee

We will now examine the relationship between $U$ and the diagonal matrices $D,K,L$ in the two cases $U = DFKFL$ and $U = DFKF^\dagger L$, respectively. We can assume the leading entries of $D$ and $L$ to be $d_{11} = l_{11} = 1$ by absorbing two phase factors in the diagonal matrix $K$.

\begin{lem}\label{tonylemma1}
Suppose $U = DFKFL$ where $D,K,L$ are diagonal unitary matrices with $D = \text{diag}(1,\alpha,\beta)$. Then either $U = PE$ where $P$ is a permutation matrix and $E$ is diagonal, or the matrix elements of $U$ are all non-zero and satisfy
\be\label{lemma1w}
u_{12}u_{23}u_{31} = u_{13}u_{21}u_{32} = u_{11}u_{22}u_{33},
\ee
and $\alpha$ and $\beta$ are given by
\be\label{lemma1alpha}
\alpha^3 = \frac{u_{21}u_{22}u_{23}}{u_{11}u_{12}u_{13}},
\ee
\be\label{lemma1beta}
\beta = \alpha^2\frac{u_{12}u_{31}}{u_{21}u_{22}}.
\ee
\end{lem}
Let $K = \text{diag}(\gamma,\delta,\epsilon)$ and $L = \text{diag}(1,\zeta,\eta)$. Then
\be\label{Wmatrix}
U = \begin{pmatrix}a & \zeta b & \eta c \\ \alpha b & \alpha\zeta c & \alpha\eta a\\\beta c & \beta\zeta a & \beta\eta b\end{pmatrix}
\ee
where
\begin{align}
a &= \third(\gamma + \delta + \epsilon),\notag\\[5pt]
b &= \third(\gamma + \omega\delta + \omega^2\epsilon),\label{abc}\\[5pt]
c &= \third(\gamma + \omega^2\delta + \omega\epsilon).\notag
\end{align}
Suppose one of $a,b,c$ were zero, say $a = 0$. Then, since $\gamma, \delta, \epsilon$ all have modulus $1$, they must form an equilateral triangle in the complex plane, so either $\delta = \omega\gamma$ and $\epsilon = \omega^2\gamma$, when $b = 0$ and $c = \gamma$, or $\delta = \omega^2\gamma$ and $\epsilon = \omega\gamma$, when $b = \gamma$ and $c = 0$. In both cases $U$ is of the form $PE$.

If none of $a,b,c$ are zero, then all the matrix elements of $U$ are non-zero and equations \eqref{lemma1w}, \eqref{lemma1alpha} and \eqref{lemma1beta} follow immediately from \eqref{Wmatrix}.

Exactly similar arguments prove

\begin{lem}\label{tonylemma2} Suppose $U = DFKF^\dagger L$ where $D,K,L$ are as in Lemma \ref{tonylemma1}. Then either $U = PE$ where $P$ is a permutation matrix and $E$ is diagonal, or the matrix elements of $U$ are all non-zero and satisfy
\be\label{lemma2w}
u_{11}u_{23}u_{32} = u_{12}u_{21}u_{33} = u_{13}u_{22}u_{31},
\ee
while $\alpha$ is given by \eqref{lemma1alpha} and $\beta$ by
\be\label{lemma2beta}
\beta = \alpha^2\frac{u_{13}u_{31}}{u_{21}u_{23}}.
\ee
\end{lem} 

We now return to eq. \eqref{W} and consider the four possibilities for $(F^{(1)},F^{(2)})$.

\noindent{\bf Case 1:} $U = DFKFL = D'FK'FL'$.

Let $D = \text{diag}(1,\alpha,\beta)$, $D' = \text{diag}(1,\alpha ', \beta')$. Then, by Lemma \ref{tonylemma1}, either $U$ is of the form $PE$ (when the bases $\mathcal{B}_0$ and $\mathcal{B}_1$ are equivalent), or
\be
\alpha^3 = \alpha'^3 \qquad \text{and} \qquad \frac{\beta'}{\beta} = \(\frac{\alpha'}{\alpha}\)^2.
\ee
Hence $\alpha' = \alpha$ or $\omega\alpha$ or $\omega^2\alpha$, so
\be
D' = \begin{pmatrix}1&0&0\\0&\alpha & 0\\0&0&\beta\end{pmatrix} \text{ or }  \begin{pmatrix}1&0&0\\0&\omega\alpha &0\\0&0&\omega^2\beta\end{pmatrix} \text{ or } \begin{pmatrix}1&0&0\\0& \omega^2\alpha &0\\0&0&\omega\beta\end{pmatrix}.
\ee
This gives
\be
V = DF = \begin{pmatrix}1&1&1\\\alpha &\omega\alpha & \omega^2\alpha\\\beta & \omega^2\beta &\omega\beta\end{pmatrix},
\ee
\be
W = D'F = \begin{pmatrix}1&1&1\\\alpha &\omega\alpha & \omega^2\alpha\\\beta & \omega^2\beta &\omega\beta\end{pmatrix} \text{ or }  
\begin{pmatrix}1&1&1\\\omega\alpha &\omega^2\alpha & \alpha\\\omega^2\beta &\omega\beta &\beta\end{pmatrix} \text{ or } 
\begin{pmatrix}1&1&1\\\omega^2\alpha & \alpha & \omega\alpha\\\omega\beta & \beta & \omega^2\beta \end{pmatrix}.
\ee
In each case the columns of $W$ are a permutation of those of $V$. Thus either the bases $\mathcal{B}_0$ and $\mathcal{B}_1$ are equivalent or $\mathcal{B}_2$ and $\mathcal{B}_3$ are equivalent.

\smallskip

\noindent{\bf Case 2:} $U = DFKFL = D'FK'F^\dagger L'$.

Suppose $U$ is not of the form $PE$. Then both Lemmas \ref{tonylemma1} and \ref{tonylemma2} apply, and $U$ has non-zero matrix elements satisfying \eqref{lemma1w} and \eqref{lemma2w}. As in case 1, let $D = \text{diag}(1,\alpha,\beta)$ and $D' = \text{diag}(1,\alpha',\beta')$. Now $\alpha$ and $\beta$ are given by Lemma \ref{tonylemma1}, but $\alpha'$ and $\beta'$ are given by Lemma \ref{tonylemma2}. Once again we have $\alpha^3 = \alpha'^3$, but now $\beta'/\beta$ is not determined solely by $\alpha'/\alpha$:
\be
\frac{\beta'}{\beta} = \(\frac{\alpha'}{\alpha}\)^2\frac{u_{13}u_{22}}{u_{12}u_{23}}.
\ee
Using \eqref{lemma1w} and \eqref{lemma2w},
\begin{align}
\( \frac{u_{13}u_{22}}{u_{12}u_{23}}\)^3 &= \(\frac{u_{13}}{u_{12}}\)^3\(\frac{u_{22}}{u_{23}}\)^3\notag\\[5pt]
&= \frac{u_{13}}{u_{12}}.\frac{u_{23}u_{31}}{u_{21}u_{32}}.\frac{u_{21}u_{33}}{u_{22}u_{31}}.\frac{u_{22}}{u_{23}}.\frac{u_{12}u_{31}}{u_{11}u_{33}}.\frac{u_{11}
u_{32}}{u_{13}u_{31}}\\[5pt]
&= 1\notag.
\end{align}
Hence $\alpha'/\alpha$ and $\beta'/\beta$ are both cube roots of $1$. Write $\alpha' = \phi\alpha$, $\beta' = \chi\beta$. If $\chi = \phi^2$ then, as shown in Case 1, the columns of $V$ and $W$ are the same, up to permutation, and the bases $\mathcal{B}_2$ and $\mathcal{B}_3$ are equivalent. If $\chi\ne\phi^2$ then two of $1,\chi,\phi$ are equal and the third is different. The same is true of the sets $\{1,\omega\chi,\omega^2\phi\}$ and $\{1,\omega^2\chi,\omega\phi\}$. Hence the sums $a = 1 + \chi + \phi$, $b = 1 + \omega\chi + \omega^2\phi$ and $c = 1 + \omega^2\chi + \omega\phi$ all have the same modulus. For $\chi\ne\phi^2$, the product
\be
V^\dagger W = F^\dagger D^\dagger D'F = \frac{1}{3}\begin{pmatrix}a&b&c\\c&a&b\\b&c&a\end{pmatrix}.
\ee
is a Hadamard matrix and hence the bases $\mathcal{B}_2$ and $\mathcal{B}_3$ are mutually unbiased. Thus in this case, $\mathcal{B}_2$ and $\mathcal{B}_3$ are either equivalent or mutually unbiased.

\smallskip

\noindent{\bf Case 3:} $U = DFKF^\dagger L = D'FK'FL'$.

This is the same as Case 2 with $V$ and $W$ interchanged.

\smallskip

\noindent{\bf Case 4:} $U = DFKF^\dagger L = D'FK'F^\dagger L'$.

This is similar to Case 1, using Lemma \ref{tonylemma2} instead of Lemma \ref{tonylemma1}. The conclusion is the same.

\smallskip

We have now shown that in every case, either $\mathcal{B}_2$ and $\mathcal{B}_3$ are equivalent or $\mathcal{B}_0$ and $\mathcal{B}_1$ are equivalent or $\mathcal{B}_2$ and $\mathcal{B}_3$ are mutually unbiased. But the assumptions of the theorem are symmetric between the pairs $\{\mathcal{B}_0,\mathcal{B}_1\}$ and $\{\mathcal{B}_2,\mathcal{B}_3\}$, so we can also prove that if $\mathcal{B}_2$ is not equivalent to $\mathcal{B}_3$ and $\mathcal{B}_0$ is not equivalent to $\mathcal{B}_1$, then $\mathcal{B}_0$ and $\mathcal{B}_1$ are mutually unbiased and therefore all four bases are mutually unbiased.


\begin{thebibliography}{30}
\bibliographystyle{plainnat}
\bibitem{ivanovic81} I. D. Ivanovi\'{c}, \textit{J. Phys. A} \textbf{14},
3241 (1981).

\bibitem{wootters+89} W. K. Wootters and B. D. Fields, \textit{Ann.
Phys. (N.Y.)} \textbf{191}, 363 (1989).

\bibitem{Durt+10} T. Durt, B-G. Englert, I. Bengtsson and K. \.{Z}yczkowski,
\textit{Int. J. Quant. Inf.} \textbf{8}, 535 (2010).

\bibitem{butterley+07} P. Butterley and W. Hall, \textit{Phys. Lett.
A} \textbf{369}, 5 (2007).

\bibitem{brierley+08} S. Brierley and S. Weigert, \textit{Phys. Rev.
A} \textbf{78}, 042312 (2008).

\bibitem{grassl+04} M. Grassl, in \textit{Proc. ERATO Conf. on Quantum
Information Science 2004 (EQIS 2004)}; also: e-print arXiv:quant-ph/0406175
(2004).

\bibitem{brierley+09} S. Brierley and S. Weigert, \textit{Phys. Rev.
A} \textbf{79}, 052316 (2009).

\bibitem{Jaming+2009} P. Jaming, M. Matolcsi, P. M\'{o}ra, F. Sz\"{o}ll\H{o}si
and M. Weiner, \textit{J. Phys. A: Math. Theor.} \textbf{42}, 245305
(2009).

\bibitem{romero+05} J. L. Romero, G. Bj\"{o}rk, A. B. Klimov, and L.
L. S\'{a}nchez-Soto, \emph{Phys. Rev. A} \textbf{72}, 062310 (2005).

\bibitem{Wiesniak+11} M. Wie\'{s}niak, T. Paterek and A. Zeilinger, \emph{New
J. Phys.} \textbf{13}, 053047 (2011).

\bibitem{Lawrence11} J. Lawrence, \emph{Phys. Rev. A} \textbf{84}, 022338 (2011).

\bibitem{bandyo+02} S. Bandyopadhyay, P. O. Boykin, V. Roychowdhury and F. Vatan, \textit{Algorithmica}, \textbf{34}, 512 (2002).

\bibitem{bennet+1999} C. H. Bennett, D. P. DiVincenzo, C. A. Fuchs,
T. Mor, E. Rains, P. W. Shor, J. A. Smolin and W. K. Wootters, \emph{Phys.
Rev. A} \textbf{59}, 1070 (1999).

\bibitem{brierley+10} S. Brierley, S. Weigert and I. Bengtsson, \textit{Quantum.
Inf. Comput.} \textbf{10}, 803 (2010).

\bibitem{bengtsson+07} I. Bengtsson, W. Bruzda, \r{A}. Ericsson, J-\r{A}.
Larsson, W. Tadej and K. \.Zyczkowski, \emph{J. Math. Phys.} {\bf 48}, 052106 (2007).

\bibitem{nielsen99} M. A. Nielsen, \textit{Phys. Rev. Lett.} \textbf{83}, 436 (1999).

\bibitem{wigner31} E. P. Wigner, \emph{Gruppentheorie und ihre Anwendung auf die Quantenmechanik der Atomspektren}, (Braunschweig, 1931).

\bibitem{klimov+07} A. B. Klimov, J. L. Romero, G. Bj\"ork and L. L. S\'anchez-Soto, \emph{J. Phys. A} \textbf{40}, 3987 (2007).

\bibitem{zauner+99} G. Zauner, \textit{Int. J. Quantum. Inf.} \textbf{9}, 445 (2011).

\bibitem{mcnulty+11} D. McNulty and S. Weigert, \emph{J. Phys. A: Math. Theor.} {\bf 45}, 102001 (2012).

\bibitem{aschbacher+07} M. Aschbacher, A. M. Childs, and P. Wocjan, \emph{J. Algebr. Comb.} \textbf{25}, 111 (2007).

\bibitem{Klappenecker+05} A. Klappenecker and M. R\"otteler,  \emph{IEEE Trans. Inform. Theory}, \textbf{51} (2005)

\end{thebibliography}
\end{document}